\documentclass[twocolumn]{aastex631}
\usepackage{lineno}

\accepted{for publication in ApJ, October 18, 2023}

\begin{document}

\title{X-ray, Near-Ultraviolet, and Optical Flares Produced By Colliding Magnetospheres in The Young High-Eccentricity Binary DQ Tau}

\correspondingauthor{Konstantin Getman}
\email{kug1@psu.edu}

\author[0000-0002-6137-8280]{Konstantin V. Getman}
\affiliation{Department of Astronomy \& Astrophysics \\
Pennsylvania State University \\ 
525 Davey Laboratory \\
University Park, PA 16802, USA}


\author[0000-0001-7157-6275]{\'{A}gnes K\'{o}sp\'{a}l}
\affiliation{Konkoly Observatory, Research Centre for Astronomy and Earth Sciences, E\"otv\"os Lor\'and Research Network (ELKH), MTA Centre of Excellence, Konkoly-Thege Mikl\'os \'ut 15-17, 1121 Budapest, Hungary}
\affiliation{Max Planck Institute for Astronomy, K\"onigstuhl 17, 69117 Heidelberg, Germany}
\affiliation{ELTE E\"otv\"os Lor\'and University, Institute of Physics, P\'azm\'any P\'eter s\'et\'any 1/A, 1117 Budapest, Hungary}

\author[0000-0003-2631-5265]{Nicole Arulanantham}
\affiliation{Space Telescope Science Institute\\ 3700 San Martin Drive, Baltimore, MD 21218, USA}

\author[0000-0002-3913-7114]{Dmitry A. Semenov}
\affiliation{Max Planck Institute for Astronomy, K\"onigstuhl 17, 69117 Heidelberg, Germany}

\author[0000-0002-6911-8686]{Grigorii V. Smirnov-Pinchukov}
\affiliation{Max Planck Institute for Astronomy, K\"onigstuhl 17, 69117 Heidelberg, Germany}

\author[0000-0002-1284-5831]{Sierk E. van Terwisga}
\affiliation{Max Planck Institute for Astronomy, K\"onigstuhl 17, 69117 Heidelberg, Germany}

\begin{abstract}
DQ~Tau is a unique young high-eccentricity binary system that  exhibits regular magnetic reconnection flares and pulsed accretion near periastron. We conducted {\it NuSTAR}, {\it Swift}, and {\it Chandra} observations during the July 30, 2022 periastron to characterize X-ray, near-ultraviolet (NUV), and optical flaring emissions. Our findings confirm the presence of X-ray super-flares accompanied by substantial NUV and optical flares, consistent with previous discoveries of periastron flares in 2010 and 2021. These observations, supported by new evidence, strongly establish the magnetosphere collision mechanism as the primary driver of magnetic energy release during DQ Tau's periastron flares. The energetics of the observed X-ray super-flares remain consistent across the three periastrons, indicating recurring energy sources during each passage, surpassing the capabilities of single stars. The observed flaring across multiple bands supports the Adams et al. model for magnetosphere interaction in eccentric binaries. Evidence from modeling and past and current observations suggests that both the mm/X-ray periastron flares and tentatively, the magnetic reconnection-related components of the optical/NUV emissions, conform to the classical solar/stellar non-thermal thick-target model, except for the distinctive magnetic energy source. However, our {\it NuSTAR} observations suffered from high background levels, hindering the detection of anticipated non-thermal hard X-rays. Furthermore, we report serendipitous discovery of X-ray super-flares occurring away from periastron, potentially associated with interacting magnetospheres. The current study is part of a broader multi-wavelength campaign, which is  planned to investigate the influence of DQ Tau's stellar radiation on gas-phase ion chemistry within its circumbinary disk.
\end{abstract}


\section{Introduction} \label{sec:intro}

DQ Tau is a nearby \citep[$D=195$~pc;][]{GaiaCollaboration2023}, non-eclipsing, double-lined spectroscopic binary system, consisting of two pre-main sequence (PMS) stars of equal mass ($0.6$~M$_{\odot}$) and equal radius ($2$~R$_{\odot}$) \citep{Mathieu1997, Czekala2016, Pouilly2023}. These stars exhibit spectral types within the range of M0 to K7. The rotational periods of the primary and secondary components are 3~days \citep{Kospal2018} and 4.5~days (Pouilly et al. in prep.), respectively. The orbital period measures 15.8 days. DQ Tau boasts a highly eccentric orbit ($e \sim 0.6$) and displays an exceptionally small periastron separation, measuring only about $8-10$ stellar radii \citep{Mathieu1997,Czekala2016, Pouilly2023}. Furthermore, the binary components of DQ Tau harbor relatively strong surface magnetic fields, estimated at around 2.5~kG, which give rise to formidable magnetospheres \citep[][Pouilly et al. in prep.]{Pouilly2023}.

Surrounding DQ Tau is a protoplanetary disk of average size ($\le 100$~au), complete with a small 0.3~au cavity \citep{Czekala2016,Kospal2018,Ballering2019}. Large optical and UV brightenings primarily occur at orbital phase ($\Phi = 0.8-1.2$), and they are mainly attributed to the pulsed accretion of disk material onto the binary components \citep{Tofflemire2017, Kospal2018, Muzerolle2019, Fiorellino2022}. However, far-UV (FUV) observations of DQ~Tau with the Cosmic Origins Spectrograph onboard {\it HST} ({\it HST}-COS) showed no correlation between the orbital phase of the binary and the C~${\rm{IV}}$ flux, a tracer of mass accretion rate, indicating that some component of the behavior is stochastic \citep{Ardila2015}.

The system exhibits powerful mm/X-ray flares coinciding with periastron passage, attributed to collisions between the magnetospheres of the binary components. The evidence supporting the magnetsophere collision hypothesis includes the recurrence of synchrotron mm-band flaring during 4 periastron encounters \citep{Salter2008,Salter2010}, the recurrence of soft X-ray flaring in 2 periastron encounters \citep{Getman2011,Getman2022b}, the timing and energy relationships between the mm and X-ray flares, and the consistency observed between the flare loop size and binary separation \citep{Salter2010,Getman2011}. 

Several other young binary systems with high eccentricities have been reported to exhibit enhanced levels of either X-ray, optical/mm, or radio emissions near their periastron passages. Notable examples include $\epsilon$~Lupi \citep{Das2023}, a collective study of four binaries (Parenago 523, RX J1622.7-2325 Nw, UZ Tau E, and HD 152404) conducted by \citet{Getman2016}, UZ Tau E \citep{Kospal2011}, and V773 Tau A \citep{Massi2008,Adams2011}. In these systems, magnetosphere collision has been proposed as a primary mechanism responsible for generating the magnetic reconnection energy that drives these events.

Modelling studies have shown that PMS X-rays have strong impact on disk ionization and chemistry \citep{Glassgold2000,Alexander2014}. But most studies assume continuous irradiation without considering the high-amplitude variations in flux and spectrum due to super-flares. A few time-dependent calculations show that disk ionization may respond to sudden large X-ray flares \citep{Rab2017,Waggoner2022}. This may already have been seen. One empirical report of ionization variability has emerged: the H$^{13}$CO$^+$ abundance of IM~Lup's disk jumped up and down by a factor of 2.5 over months \citep{Cleeves2017}. However, since no concurrent X-ray observations were conducted, the exact cause cannot be definitively ascribed to X-ray flaring.

The occurrence of large X-ray flares in young stars is relatively rare and unpredictable, displaying a stochastic nature \citep{Getman2021}. However, the presence of predictable X-ray super-flares and accretion outbursts in close proximity to periastron passage makes DQ Tau an extraordinary laboratory for investigating the influence of stellar radiation on the gas-phase ion chemistry within its disk.

As part of our comprehensive multi-wavelength research program, dedicated to examining the effects of UV and X-ray radiation on the disk of DQ~Tau, our team was awarded valuable ALMA observation time. Specifically, our aim was to observe the response of H$^{13}$CO$^{+}$ emission throughout a single orbit, as a reaction to an increase in ionizing radiation in the vicinity of a related periastron passage.

To achieve this, we conducted a single X-ray observation using the {\it NuSTAR} telescope, along with multiple short-duration observations using the {\it Swift} telescope in the X-ray, UV, and optical wavelengths. These observations were strategically timed near a specific periastron passage of DQ~Tau, taking place in July-August 2022. Additionally, we captured multiple snapshots of the non-periastron portion of DQ~Tau's orbit using the {\it Chandra} X-ray telescope. However, we encountered unfavorable weather conditions that limited our ALMA observations to a single short session near periastron. The results from this ALMA observation will be presented in a forthcoming paper.

Meanwhile, building upon the X-ray/UV/optical data acquired in 2022 and the previously obtained X-ray/mm data \citep{Salter2010,Getman2011,Getman2022b}, our current study is dedicated to further investigating the origins and energetic properties of X-ray flares, along with their corresponding near-ultraviolet (NUV) and optical flare counterparts, observed within the remarkable young binary system DQ~Tau.

The structure of the paper is outlined as follows: Section~\ref{sec:xray_data_reduction} provides a detailed description of the data reduction and reduction procedures employed for X-ray, UV, and optical analyses. In Section~\ref{sec:xray_flare_analyses}, we present the detection of flares and examine their spectral properties. Section~\ref{sec:comparison_with_coup_mystix_sfincs} offers a comparison between the X-ray periastron flares observed in DQ Tau and super-flares observed in numerous other PMS stars. Finally, Section~\ref{sec:discussion} delves into a discussion surrounding the origin and energetics of the X-ray flares, as well as their associated NUV and optical counterparts.

\begin{figure*}
\centering
\includegraphics[width=0.95\textwidth]{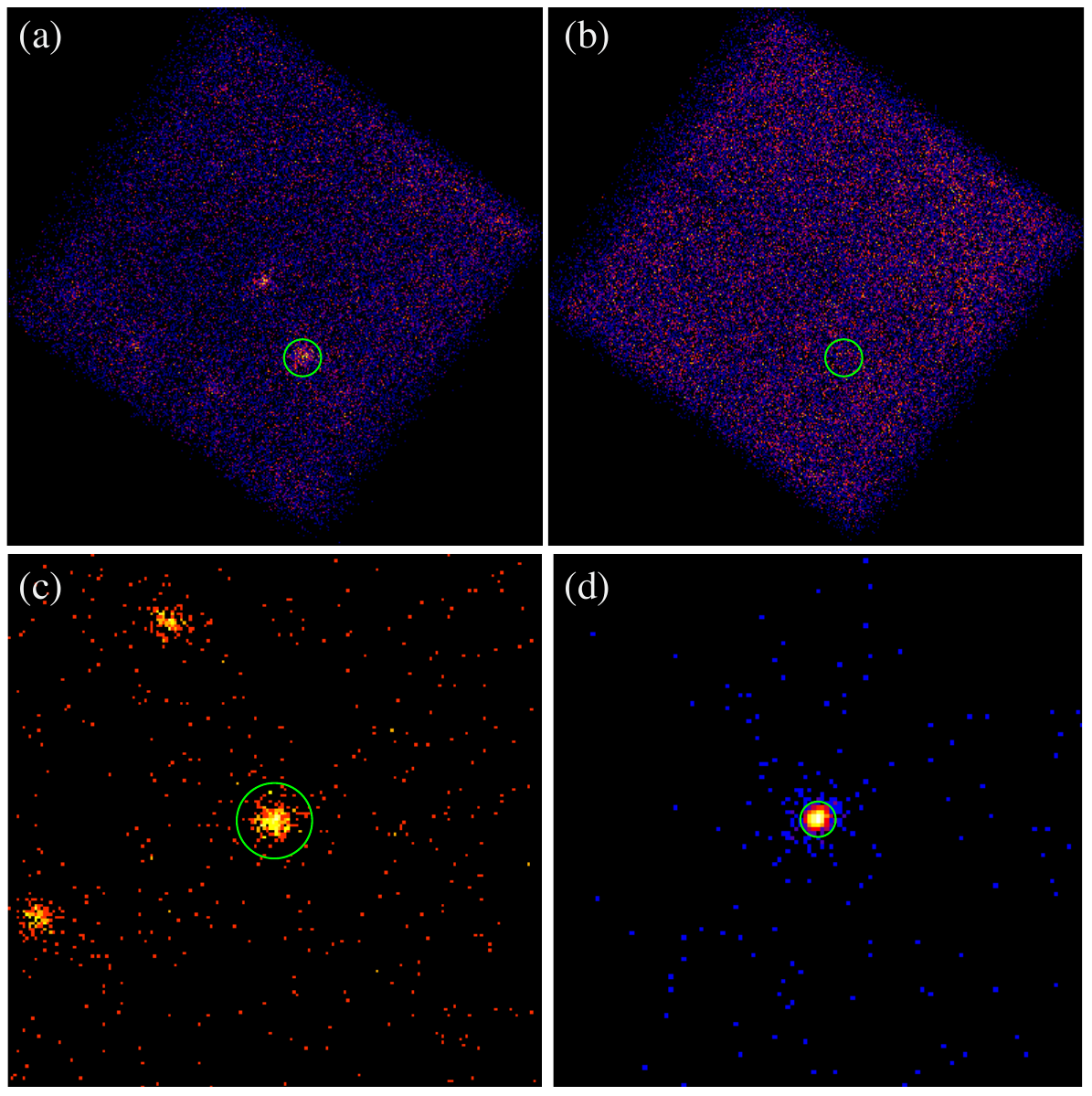}
\caption{{\it NuSTAR}, {\it Swift}, and {\it Chandra} images of DQ~Tau and its neighborhood.  The color map, ranging from yellow to red to blue, indicates the intensity of X-rays with yellow and blue pixels representing more and less X-ray counts, respectively. Each figure panel has its own unique intensity scale. DQ~Tau source extraction regions are marked by the green circles. (a) $13\arcmin \times 13\arcmin$ image from the merged {\it NuSTAR} FPMA$+$FPMB event lists in the ($3-10$) keV energy band, and (b) in the ($10-50$)~keV energy band. (c) $7\arcmin \times 7\arcmin$ {\it Swift}-XRT cutout of DQ~Tau's neighborhood in the (0.2-10)~keV band, which was obtained from the merged event lists of the 16 {\it Swift} observations. The other two X-ray objects visible in this image are the young stellar systems Haro~6-37 A,B and DR~Tau, located to the north-east and south-east of DQ~Tau, respectively. (d) $1\arcmin \times 1\arcmin$ {\it Chandra}-ACIS-I3 cutout in the ($0.5-8$) keV band centered on DQ~Tau. This image was obtained from the merged event lists of the 12 {\it Chandra} observations.} \label{fig:xray_images}
\end{figure*}

\section{X-ray Observations and Data Extraction} \label{sec:xray_data_reduction}

\subsection{{\it NuSTAR} Data} \label{sec:nustar_data}

We conducted a single observation (ObsID 30801011002; PI K.~Getman) of this system with {\it NuSTAR} \citep{Harrison2013} from 04:51:09 UTC on July 28, 2022, to 19:36:09 UTC on July 31, 2022. This time window covers the orbital phase range of $\Phi=(0.9 - 1.13)$ around the periastron passage on July 30, 2022 ($\Phi=1.0$). The science exposure of the observation, taking into account Earth occultations, is 156 ksec.

The data obtained from focal plane modules A and B (FPMA and FPMB), covering similar energy ranges, were processed using tools from the {\it NuSTAR} Data Analysis Software package {\it NuSTARDAS} (v. 2.1.2), which is incorporated into {\it HEASOFT} (v. 6.31.1) as detailed in \citep{Heasoft2014}. {\it NuSTAR}'s calibration database (CALDB) v. 20221229 was utilized. The data went through calibration and screening using the {\it nupipeline} tool. Parameters saamode=OPTIMIZED and tentacle=yes were employed to screen the data for elevated count rates resulting from the spacecraft's passages through the South Atlantic Anomaly. Subsequently, the {\it nuproducts} tool was utilized to generate various outputs, including the source and background lightcurves and spectra, as well as response and response matrix files.

Figures~\ref{fig:xray_images}(a,b) present images of the combined FPMA and FPMB event lists for the $(3-10)$~keV and $(10-50)$~keV energy bands, respectively. The source counts were obtained from a circular region with a radius of $40\arcsec$ (indicated by the green circle in Figure~\ref{fig:xray_images}), representing 60\% of the energy within the point spread function. The background measurement was performed locally in an area devoid of sources. Within the source extraction circle, there are 771 X-ray events with energies ranging from 3 to 10~keV, approximately one third of which are background events. The source is not detected in the $(10-20)$~keV band (not shown) or the $(10-50)$~keV band (Figure~\ref{fig:xray_images}b), nor at higher energies.

\subsection{{\it Swift} Data} \label{sec:swift_data}
Using the Neil Gehrels \textit{Swift} observatory \citep{Gehrels2004}, we conducted 16 short observations of DQ Tau near periastron over a time period of July 28 to August 2, 2022. These observations are part of our joint {\it NuSTAR}/{\it Swift} program. The observations were spaced several hours apart, with durations ranging from 1 to 1.7~ksec, totaling 22.5~ksec. The target ID for these observations is 14857. The X-ray Telescope (XRT) operated in the PC mode, while the Ultraviolet/Optical Telescope (UVOT) operated in the 0x30ed standard six-filter blue-weighted mode. The {\it Swift}-XRT data product generator \citep{Evans2007,Evans2009} was utilized to construct X-ray light curves and source/background spectra, along with the relevant calibration files. The generator employed {\it HEASOFT} package (v. 6.29) and CALDB (v. 20230109).

Figure \ref{fig:xray_images}c presents the XRT image obtained from merging the event lists of all 16 observations. Within the circular source extraction region of a $30\arcsec$ radius (indicated by the green circle in the image), we identified 249 X-ray counts with energies ranging from 0.2 to 10 keV, of which only a few percent represented background counts.

For each of the 16 observations, UVOT magnitudes for the six filters (V:B:U:W1:M2:W2) were measured by applying the {\it fappend} \citep{Blackburn1999} and {\it uvotmaghist} tools from {\it HEASOFT} (v. 6.31.1).

\subsection{{\it Chandra} Data} \label{sec:chandra_data}

To investigate the soft X-ray emission throughout the entire orbital phase of DQ Tau and complement the observations made by \textit{NuSTAR} and \textit{Swift} at periastron, additional X-ray data were obtained utilizing the {\it Chandra X-ray Observatory} \citep{Weisskopf2002}. The investigation involved 12 short \textit{Chandra} imaging observations of DQ Tau away from periastron, with each observation lasting approximately 1.5~ksec. These observations were part of the Director's Discretionary Time (DDT) program, with corresponding observation IDs ranging from 26464 to 26475.

Data were obtained between August 1 and August 14, 2022, covering an orbital phase range of 1.1 to 1.9. To mitigate potential pileup effects during anticipated X-ray flares, a 1/8 sub-array of a single ACIS-I3 chip was employed \citep{Garmire2003}.

For the {\it Chandra} data reduction and analysis, CIAO v4.15\citep{Fruscione2006} and CALDB v4.10.4 were utilized. The CIAO tools {\it chandra\_repro} and {\it reproject\_obs} were used to reprocess the data and merge the event images. Figure \ref{fig:xray_images}d displays the cutout of the merged {\it Chandra}-ACIS-I3 image of DQ~Tau. Count rates and apparent fluxes were measured, and spectra and response files were generated using the \textit{srcflux} tool. Within the circular source extraction region, with a radius of $2\arcsec$ (indicated by the green circle in the image), and within the energy band of $(0.5-8)$ keV, net X-ray events were observed between 30 and 50 per observation in 8 instances. However, the remaining 4 consecutive observations (ObsIDs 26471, 26472, 26473, and 26474) exhibited higher net count levels, ranging from 61 to 781 counts per observation.

Despite using the 1/8 sub-array ACIS mode with a reduced CCD frame of 0.5 seconds, data from one of the observations with the highest count rate (ObsID 26471) suffered from pileup. Assuming a single-temperature optically thin thermal plasma with $kT \sim 2$~keV and an average X-ray column density of $N_H \sim 1.3 \times 10^{21}$~cm$^{-2}$ \citep{Getman2011,Getman2022b}, the Portable, Interactive Multi-Mission Simulator (PIMMS) estimated a pileup fraction (the ratio of the number of frames with two or more events to the number of frames with one or more events) of 13\%, resulting in an increase in the apparent count rate of up to 30\%.

\begin{figure*}
\centering
\includegraphics[width=0.95\textwidth]{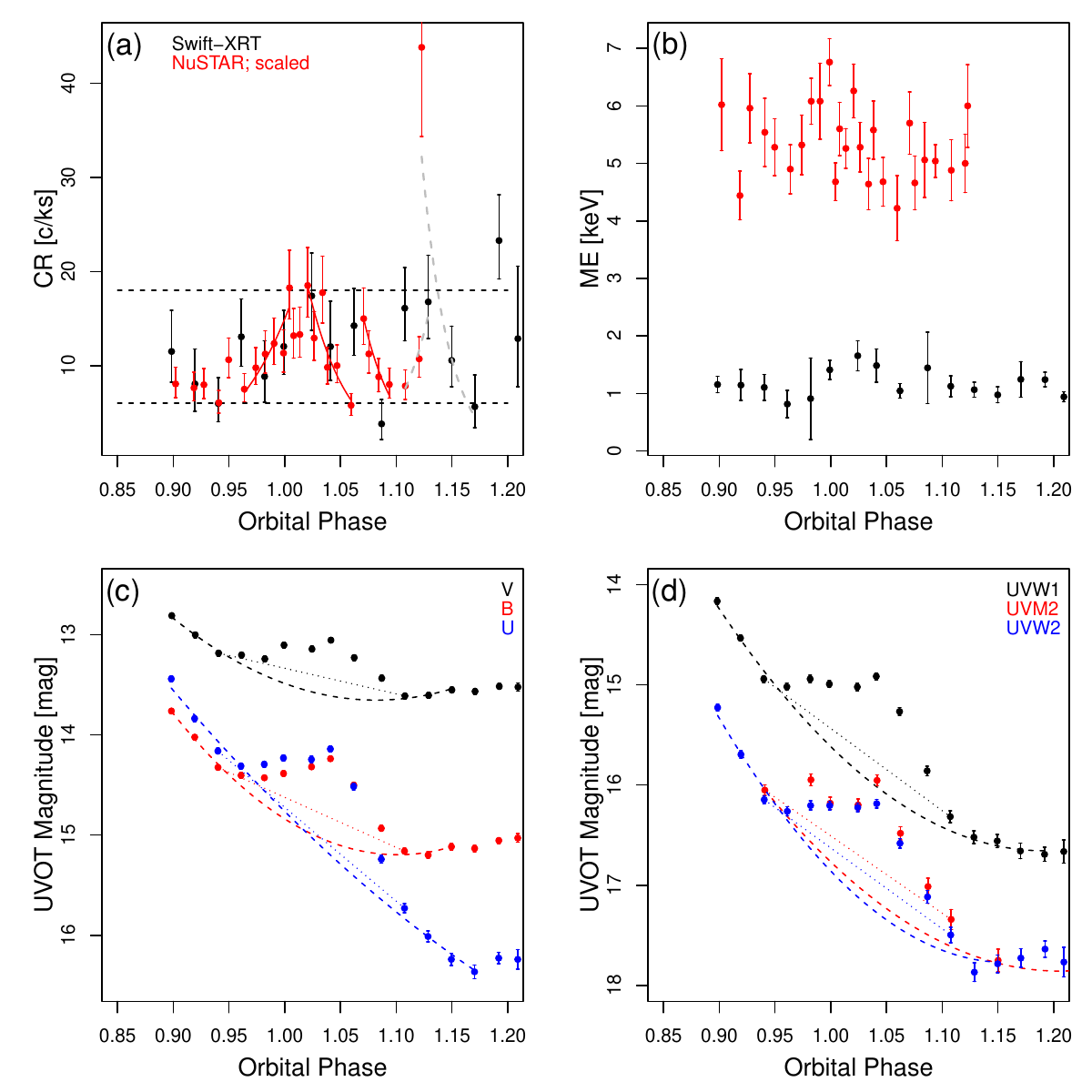}
\caption{{\it NuSTAR} and {\it Swift} lightcurves and related temporal evolution of the X-ray median energy. The error bars indicate the 68\% confidence intervals for all the shown quantities. (a) The background-subtracted lightcurve for the combined FPMA and FPMB {\it NuSTAR} data is shown in red. The solid red curves show the best-fit exponential fits to the rise and decay phases of the main flare, and to the decay phase of the second flare. The dashed grey curves indicate exponential fits for the rise and decay phases of the third flare. The {\it Swift}-XRT data, with each point corresponding to a single observation, are displayed in black. The {\it NuSTAR} count rate is scaled by a factor of $\times 3$ to match the {\it Swift} count rate level. The average lowest {\it NuSTAR}+{\it Swift}-XRT count rate level across the entire observed periastron phase and the peak level of the main flare are marked by the black dashed lines. (b) The temporal evolution of the X-ray median energy is shown for both the {\it NuSTAR} (red) and {\it Swift} (black) data. (c,d) {\it Swift}-UVOT lightcurves are provided for the six UVOT filters. Polynomial and linear fits to the initial and final segments of the flare light curves, created using the R function \textit{lm}, are depicted by the dashed and dotted colored lines. These fits serve as potential baselines for the UVOT flaring resulting from magnetic reconnection.} \label{fig:nustar_swift_lcs}
\end{figure*}

\begin{figure*}
\centering
\includegraphics[width=0.95\textwidth]{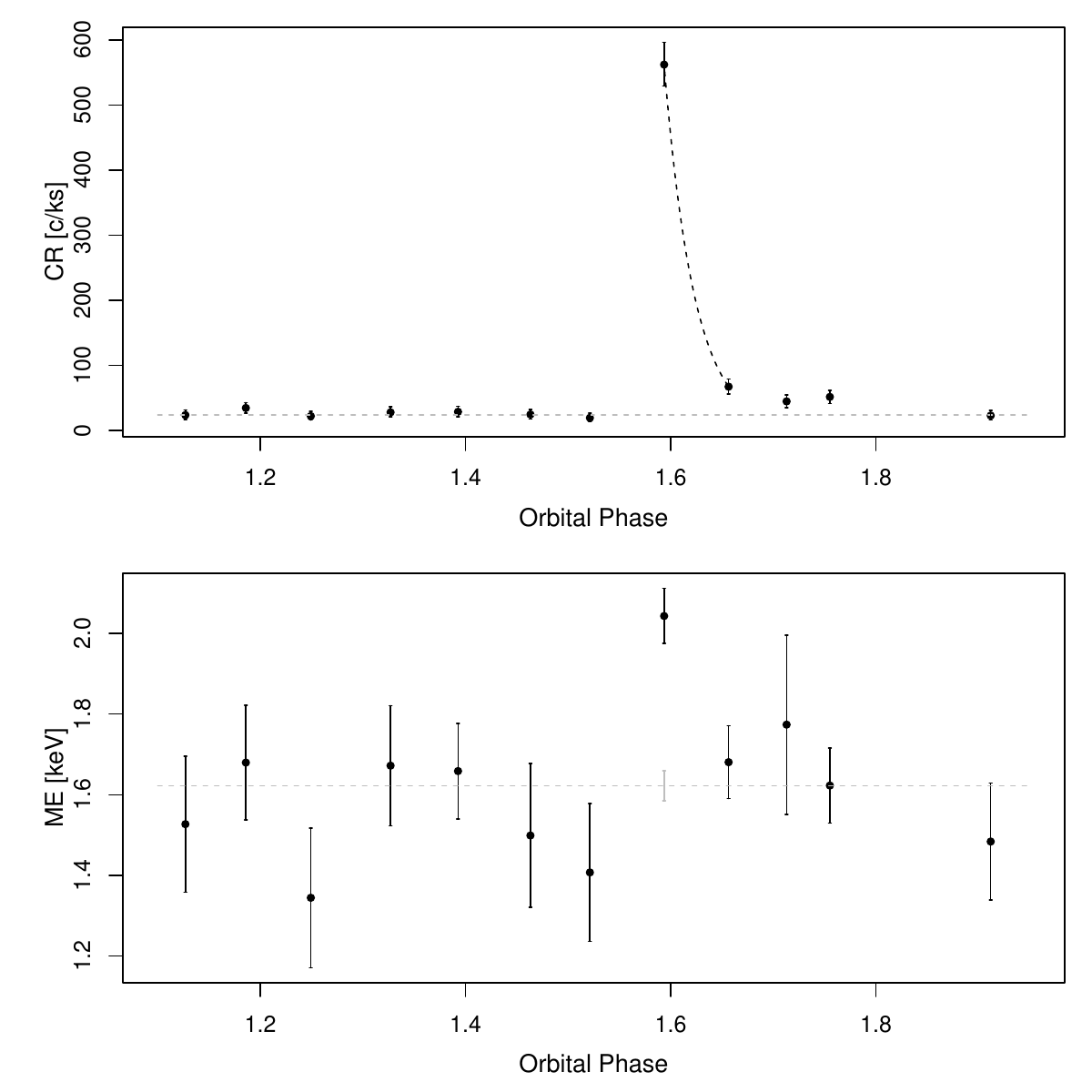}
\caption{The {\it Chandra} lightcurve and the related temporal evolution of the X-ray median energy. Each point represents a single {\it Chandra} observation, and the error bars indicate the 68\% confidence intervals for all the shown quantities. (a) On the shown lightcurve, the dashed grey line marks the characteristic level of the X-ray emission in DQ~Tau. The point with the highest count rate corresponds to the observation \# 26471, and this count rate value is not corrected for pileup. The dashed black line connects the two points corresponding to the observations \#\# 26471 and 26472 and indicates the exponential decay of the X-ray emission assuming that these points are associated with the same large X-ray flare. (b) The evolution of the X-ray median energy is displayed. The dashed grey line and grey error bar mark the median level of the energy and its 68\% uncertainty for the data comprising all but one observation (\# 26471 with the highest count rate and median energy values).} \label{fig:chandra_lc}
\end{figure*}

\section{Flare Analyses} \label{sec:xray_flare_analyses}

\subsection{Detection of Flares} \label{sec:xray_lcs}

Figures~\ref{fig:nustar_swift_lcs}a,b and \ref{fig:chandra_lc}a,b display the X-ray lightcurves and temporal evolution of the median energy for the X-ray events detected within the circular extraction regions depicted in Figure~\ref{fig:xray_images}.

In the shown lightcurve and median energy plots for {\it NuSTAR}, individual data points represent bins containing 30 X-ray events from the combined FPMA and FPMB data. Conversely, each point in the {\it Swift}-XRT and {\it Chandra} lightcurves and median energy plots corresponds to a single X-ray observation.

The X-ray lightcurve from {\it NuSTAR} provides clear evidence of the occurrence of at least two X-ray flaring events within the orbital phase range of $(0.96-1.1)$. The first (main) flare is identified by solid red curves representing exponential fits to the rise and decay phases, while the second flare is indicated by the decay data fit. These fits were performed using the observed binned count rate data, as described in equation (B1) and detailed in \citet{Getman2021b}. The resulting time scales are as follows: an $80 \pm 19$~ksec rise time ($\tau_{rise}$) for the main flare and decay times ($\tau_{decay}$) of $49 \pm 12$~ksec and $52 \pm 13$~ksec for the main and second flares, respectively. Such decay time scales are commonly observed in large X-ray flares detected in numerous young stellar members of various nearby star-forming regions \citep{Getman08a,Getman2021}. However, the main flare's long rise time is longer than that of typical large X-ray flares, reminiscent of rare slow-rise-top-flat flares observed in a dozen young stellar members of the Orion Nebula Cluster \citep{Getman08a}.

The approximate observed peak level of the main X-ray flare is indicated by the upper dashed line in Figure~\ref{fig:nustar_swift_lcs}. It surpasses the "characteristic" level (baseline) of X-ray emission \citep{Wolk05,Caramazza07}, represented by the lower dashed line, by a factor of 3. The characteristic level likely corresponds to the combined effect of numerous unresolved micro-flares and nano-flares. The decay time-scale and amplitude of the main flare bear resemblance to those measured for the large X-ray flares captured near DQ~Tau's periastron by {\it Chandra} and {\it Swift}-XRT in 2010 and 2021 \citep{Getman2011,Getman2022b}.

The {\it Swift}-XRT data, despite lower cadence and counting statistics, exhibit a similar morphology to the {\it NuSTAR} DQ~Tau flares. Moreover, both the X-ray median energies from {\it NuSTAR} and {\it Swift}-XRT display temporal evolution patterns of rise and decay within the orbital phase range of $(0.96-1.1)$. The median energy serves as a proxy for plasma temperature, and these temperature evolutionary patterns are characteristic of X-ray flares fueled by magnetic reconnection processes \citep{Getman08a,Getman2011,Getman2021b}.

Furthermore, the {\it NuSTAR} and {\it Swift}-XRT light curves suggest the presence of additional X-ray flaring events occurring beyond the orbital phase of 1.1. We postulate the existence of at least two significant X-ray flares within the orbital phase intervals of $(1.1-1.17)$ and $(1.17-1.21)$. These events are designated as the third and fourth X-ray flares, respectively. The third flare may comprise smaller flares within it. Our estimates for the rise and decay timescales of the third flare, derived from a combination of {\it NuSTAR} and {\it Swift}-XRT data points, are $\tau_{rise} = 40 \pm 10$~ks and $\tau_{decay} = 33 \pm 11$~ks, as indicated by the grey dashed lines in Figure~\ref{fig:nustar_swift_lcs}a. These measurements are only approximate due to differences between the {\it NuSTAR} and {\it Swift}-XRT data points near orbital phases 1.11 and 1.13, which may suggest a more complex flare morphology.   Unfortunately, we lack sufficient {\it Swift}-XRT data to determine the timescales for the fourth flare. 

Figures~\ref{fig:nustar_swift_lcs}c,d illustrate the {\it Swift}-UVOT lightcurves for the six UVOT filters. While the longer time-scale UVOT variation may be linked to the known increased accretion rate of material from the circumbinary disk onto both stellar components during periastron passages \citep[][and references therein]{Fiorellino2022}, the shorter time-scale variation, occurring within the $(0.96-1.1)$ orbital phase range, is associated with the main and second X-ray flares. The classical non-thermal thick-target model, frequently applied to solar and stellar flares \citep{Brown1971, Lin1976}, predicts an optical component to accompany an X-ray flare. The simultaneous appearance of the {\it Swift}-UVOT and {\it NuSTAR}+{\it Swift}-XRT brightening events is qualitatively in line with the model predictions and suggests that both X-ray and UVOT emissions trace the same astrophysical phenomenon. It also aligns with empirical data on optical-X-ray flares observed in young members of the NGC~2264 star-forming region \citep{Flaccomio2018}. 


Noticeably absent are UVOT flaring counterparts corresponding to the third and fourth X-ray flares in DQ~Tau. This observation aligns with the prevailing hypothesis that the UVOT flaring emissions originate in close proximity to the stellar surface, near the foot-points of the extended X-ray flaring structures. Consequently, these emissions can occasionally be obscured by the stellar limbs of both DQ~Tau stellar components, as discussed in \citet{Flaccomio2018}. It  is worth recalling that the rotation periods of both stellar components, which fall in the range of $3-4.5$~days, are comparable to the combined duration of the first two X-ray flares. As noted by \citet{Flaccomio2018}, a similar phenomenon was observed in their study of young stars in NGC~2264, where approximately 50\% of the large X-ray flares lacked optical flaring counterparts for the same underlying reason.

Figure~\ref{fig:chandra_lc} showcases our {\it Chandra} observation made at orbital phases away from the periastron, capturing a potent X-ray flare. The observed peak count rate (ObsID 26471; orbital phase 1.60), uncorrected for pileup effects, exceeds the characteristic count rate level (represented by the grey dashed line) by a factor greater than 20. If the {\it Chandra} data point with the second highest count rate (ObsID 26472; orbital phase 1.66) is linked to the same flare, then the exponential decay time scale (marked by the black dashed line) may extend up to 40~ksec. Notably, the X-ray median energy at the observed peak of the flare, as measured from the ObsID 24671 data, significantly surpasses the values obtained from the combined data of the other 11 {\it Chandra} observations (illustrated by the grey dashed line with the accompanying grey solid error bar), indicating a hotter plasma state during the flare. The conclusive evidence for this finding will be provided by the {\it Chandra} pileup-corrected spectroscopy, which will be discussed in detail in \S\ref{sec:xray_spectra}.

\begin{figure*}
\centering
\includegraphics[width=0.95\textwidth]{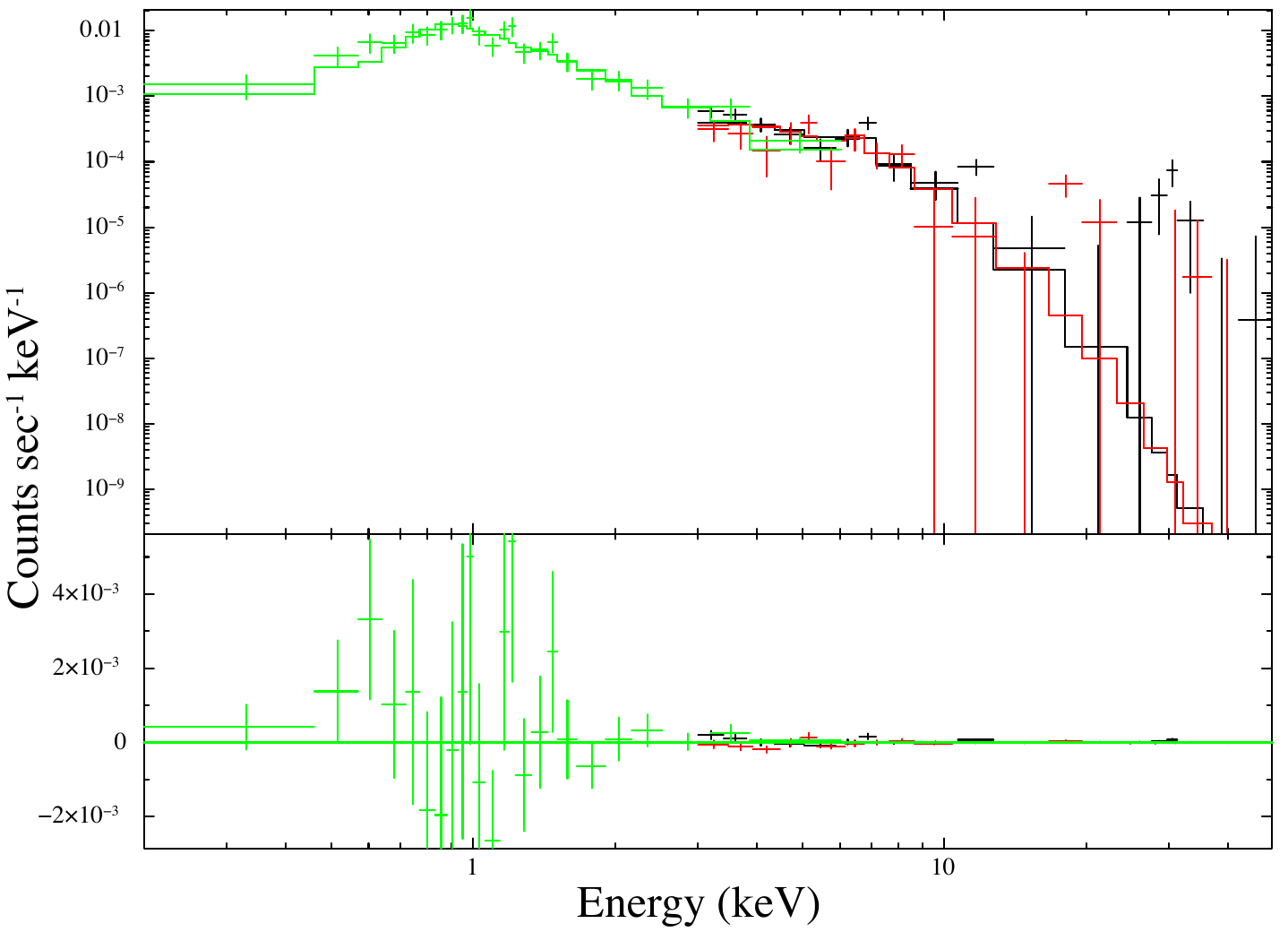}
\caption{Simultaneous fitting of the {\it Swift} and {\it NuSTAR} X-ray spectra. The {\it Swift}-XRT spectrum (shown in green) is merged from the data of all 16 {\it Swift} snapshots covering the $(0.9-1.22)$ orbital phase range. The {\it NuSTAR} FPMA (black) and FPMB (red) spectra are from the data of the entire {\it NuSTAR} observation that spans the $(0.9-1.15)$ phase range. The best-fit optically thin thermal plasma model is shown with the green, black, and red solid lines. The bottom figure panel shows the residuals between the data and the model.} \label{fig:nustar_swift_spectra}
\end{figure*}

\newpage
\begin{deluxetable*}{cccccc}
\tabletypesize{\normalsize}
\tablecaption{X-ray Spectroscopy \label{tab:xray_spectral_results}}
\tablewidth{0pt}
\tablehead{
\colhead{Spectrum} & \colhead{$\chi^{2}_{\nu}$} & \colhead{$dof$} &
\colhead{$kT_2$} & \colhead{$EM_2/EM_1$} & \colhead{$L_X$} \\
\colhead{} & \colhead{} & \colhead{} &  \colhead{(keV)} & \colhead{} & \colhead{($10^{30}$~erg~s$^{-1}$)}\\
\colhead{(1)} & \colhead{(2)} & \colhead{(3)} & \colhead{(4)} & \colhead{(5)} & \colhead{(6)}
}
\startdata
{\it Swift}$+${\it NuSTAR} & 1.2 & 57 & $3.0\pm0.4$ & $1.8\pm0.5$ & $\sim 2$; $\sim 6$\\
{\it Chandra} flare & 0.8 & 42 & $5.4\pm1.2$ & $5.4\pm1.5$ & $52$\\
{\it Chandra} characteristic$+$post-flare & 1.3 & 41 & $2.6\pm0.6$ & $0.8\pm0.2$ & $2$\\
\enddata 
\tablecomments{Column 1: Spectral data. The first row is associated with Figure~\ref{fig:nustar_swift_spectra}; the next two rows are associated with Figure~\ref{fig:chandra_spectrum}. Columns 2-3: Reduced $\chi^{2}$ for the overall spectral fit and degrees of freedom. Column 4:  Inferred
temperature of the hot plasma component and its 1 $\sigma$ error. Column 5: Inferred ratio of the emission measures for the hot and cool plasma components, and its 1 $\sigma$ error. Column 6: Inferred X-ray luminosity in the $(0.5-8.0)$~keV band. In the first row, the listed X-ray luminosity values were obtained using the {\it Swift} spectral component and are for the characteristic X-ray level and the peak of the main periastron X-ray flare at the orbital phase $\sim 1$ (Figure~\ref{fig:nustar_swift_lcs}a). These values are preceded by the ``$\sim$'' sign to indicate that the scaling from count rates to X-ray luminosities, using the time-integrated correction factor, is approximate, as it does not account for the temporal evolution of the X-ray emission hardness. In the second  row, the listed X-ray luminosity is for the observed peak of the {\it Chandra} flare at the orbital phase $1.6$ (Figure~\ref{fig:chandra_lc}a); this value is corrected for pileup by including the multiplicative {\it XSPEC} model component {\it pileup}. In the third row, the listed X-ray luminosity is for the characteristic emission level (dashed grey line in Figure~\ref{fig:chandra_lc}a); this value was corrected for the contribution of the post-flare data included in the spectrum.}
\end{deluxetable*}
\newpage

\begin{figure*}
\centering
\includegraphics[width=0.8\textwidth]{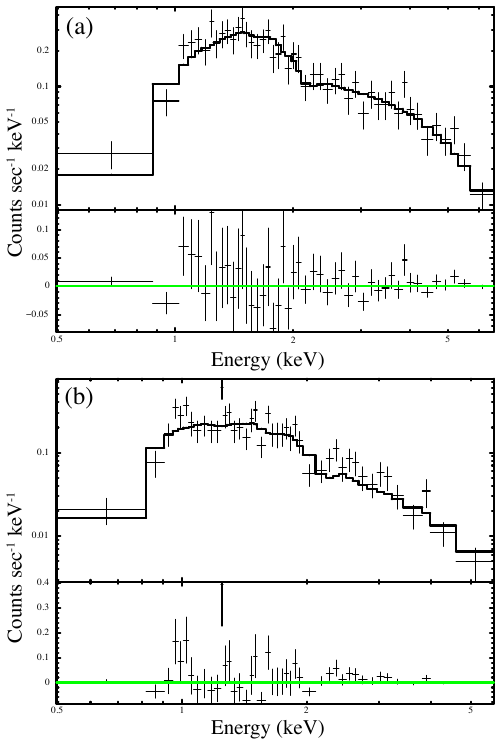}
\caption{Fitting of the {\it Chandra} spectra. The underlying {\it Chandra} data cover the $(1.1-1.9)$ orbital phase range. (a) The ``flare'' spectrum at the orbital phase $1.6$ obtained from the single snapshot observation \# 26471. (b) This spectrum is merged from the data of the remaining 11 {\it Chandra} snapshots that include the characteristic (8 observations; Figure~\ref{fig:chandra_lc}) and post-flare (3 observations) X-ray emission states. The best-fit optically thin thermal plasma models are shown with the black solid lines. The bottom sub-panels of each figure panel show the residuals between the data and the model.} \label{fig:chandra_spectrum}
\end{figure*}
\newpage

\subsection{Spectral Analyses Of Flares and Characteristic Emission} \label{sec:xray_spectra}

The {\it NuSTAR} and {\it Swift}-XRT data cover a similar orbital phase range, but they provide only modest counting statistics. Consequently, only basic spectral properties, such as the time-integrated hot plasma temperature component and X-ray luminosities (specifically for the characteristic and peak main flare states), can be reliably determined\footnote{The conversion factor derived from this spectral analysis, which relates the observed time-integrated count rate to the intrinsic X-ray luminosity, is further applied to scale the count rates and calculate the intrinsic X-ray luminosities for both the characteristic and flare peak states.}. To achieve this, we performed simultaneous fitting of the stacked {\it Swift}-XRT spectrum from the 16 observations and the individual {\it NuSTAR} FPMA and FPMB spectra. The {\it Swift}-XRT spectrum was binned to a minimum of 10 counts per bin, whereas the {\it NuSTAR} spectra were binned to 40 counts per bin. We employed the simple absorbed two-temperature optically thin thermal plasma model using the {\it XSPEC} package \citep{Arnaud1996} and employed Gehrels $\chi^{2}$ statistics \citep{Gehrels86} for the data fitting. The model used was $tbabs \times (apec + apec)$, where $tbabs$ \citep{Wilms2000} and $apec$ \citep{Smith2001} represent the absorption and plasma emission model components, respectively.

Considering the low counting statistics, certain parameters were held fixed at characteristic values. Specifically, the coronal elemental abundances, the soft temperature component ($kT_{1}$), and the column density ($N_H$) were all set to their respective characteristic values. The coronal elemental abundances were fixed at 0.3 times solar elemental abundances \citep[][for young stars]{Imanishi2001,Feigelson2002}, the soft temperature component ($kT_{1}$) was set to $0.8$~keV \citep[][for young stars]{Preibisch05,Getman10}, and the column density ($N_H$) was held at $1.3 \times 10^{21}$cm$^{-2}$ \citep[][for DQ~Tau]{Getman2011}.

Figure~\ref{fig:nustar_swift_spectra} showcases the {\it Swift}-XRT and {\it NuSTAR} spectra along with the best-fit model obtained from the simultaneous fit. The corresponding spectral fitting results can be found in Table~\ref{tab:xray_spectral_results}. To determine the X-ray luminosities, the count rate of the {\it Swift}-XRT spectrum was scaled to the count rate levels corresponding to the peak of the main flare and characteristic states shown in Figure~\ref{fig:nustar_swift_lcs} (dashed lines).

Furthermore, the inclusion of the {\it powerlaw} model component to search for non-thermal X-ray emission did not yield an improved fit, confirming the findings of Figure~\ref{fig:xray_images}b, where it was evident that the {\it NuSTAR} data beyond 10~keV were predominantly influenced by the background.

In contrast to the {\it Swift}-XRT and {\it NuSTAR} data, the {\it Chandra} data provide a sufficient number of X-ray counts, allowing for the creation of two distinct spectra. One spectrum is derived from the data of a single observation (ObsID 26471) characterized by the highest count rate, representing the observed flare peak. The second spectrum is based on the data from the remaining 11 {\it Chandra} observations, corresponding to the characteristic and post-flare states. To ensure reliable analysis, the "observed flare peak" and "characteristic $+$ post-flare" spectra were binned, with a minimum of 15 counts per bin for the former and 10 counts per bin for the latter.

The {\it Chandra} "characteristic $+$ post-flare" spectrum was fitted using the same model and fixed parameters employed for the {\it NuSTAR} and {\it Swift}-XRT spectra. However, for the flare peak spectrum, as the ObsID 26471 data are affected by pileup, the $pileup$ model described by \citet{Davis2001} was applied, along with $tbabs \times (apec + apec)$, to correct for this effect.  Only the grade morphing  $\alpha$ and PSF fraction $psffrac$ parameters were varied within the {\it pileup} model. The best-fit parameters are $\alpha = 0.9$ and $psffrac = 0.98$. The spectral fitting of {\it Chandra} data is illustrated in Figure~\ref{fig:chandra_spectrum}, with the corresponding results summarized in Table~\ref{tab:xray_spectral_results}.

The X-ray spectral analyses described above yield the following key findings. Firstly, the {\it NuSTAR} data do not reveal any evidence of non-thermal X-ray emission near or around the July 30, 2022 periastron. Secondly, the inferred characteristic (baseline) X-ray luminosity of $L_X = 2 \times 10^{30}$~erg~s$^{-1}$ during the July-August, 2022 orbit of DQ~Tau aligns well with DQ~Tau's X-ray baselines observed during the periastron passages in 2010 and 2021 \citep{Getman2022b}. Thirdly, the inferred temperature of the hot plasma component ($kT_{2}$) and the ratio of emission measures for the hot and cool plasma components ($EM_2/EM_1$) indicate that the coronal gas is hottest during the observed peak of the {\it Chandra} flare away from periastron, warm near periastron, and coolest during the characteristic and post-flare states away from periastron. Fourthly, the X-ray luminosity at the observed peak of the non-periastron flare is 8.6 times higher than that at the observed peak of the main periastron flare.

\subsection{Flare Energetics and Loop Length} \label{sec:flare_energy_loop_length}

\subsubsection{X-ray Energy Of Periastron Flares} \label{sec:flare_energy_xray}

We have estimated the energy values of three X-ray flares occurring near periastron, referred to as the main, second, and third flares. This estimation involves integrating the count rates from {\it NuSTAR} and {\it Swift}-XRT, as shown in Figure~\ref{fig:nustar_swift_lcs}a. These counts were corrected for the baseline and transformed into intrinsic X-ray luminosities, all within the duration of the respective flares. We excluded the outlier {\it Swift}-XRT point at orbital phase 1.11 from the calculation of the third flare's energy, as it may be related to a smaller unresolved flare.

The resulting energies for these flares are as follows. Main Flare: $E_{X,main} = (3.2 \pm 2.3) \times 10^{35}$~erg. Second Flare: $E_{X,second} = (1.1 \pm 0.8) \times 10^{35}$~erg. Third Flare: $E_{X,third} = (2.0 \pm 0.5) \times 10^{35}$~erg. When considering the combined X-ray flare energetics in the vicinity of periastron, within the orbital intervals $\Phi=(0.96 - 1.1)$ and $\Phi=(0.96 - 1.17)$, the total energies are $E_X = (4.3 \pm 2.4) \times 10^{35}$~erg and $E_X = (6.3 \pm 2.5) \times 10^{35}$~erg, respectively. All the energies were calculated within the $(0.5-8)$~keV band.

For the fourth periastron X-ray flare, the three {\it Swift}-XRT points suggest energy levels comparable to those of the first three flares. However, the limited data points available prevent us from performing a more sophisticated energy calculation for this flare.

\subsubsection{Optical and Near-Ultraviolet Energies Of Periastron Flares} \label{sec:flare_energy_optical}

Here, we categorize the {\it Swift}-UVOT observations into two groups: optical and near-ultraviolet (NUV). The optical data correspond to measurements taken with the $V$, $B$, and $U$ filters, which have central wavelengths of $547$~nm, $439$~nm, and $346$~nm, respectively. On the other hand, the NUV data correspond to observations made with the $W_1$, $M_2$, and $W_2$ filters, which have central wavelengths of $260$~nm, $225$~nm, and $193$~nm, respectively. Note that several {\it Swift}-UVOT observations lack data in the $M_2$ filter (Figure~\ref{fig:nustar_swift_lcs}), resulting in fewer NUV data points than optical points used in the analyses below.

Figure~\ref{fig:nustar_swift_lcs} illustrates that the main and second X-ray periastron flares detected by {\it NuSTAR} and {\it Swift}-XRT are accompanied by significant optical and NUV flares detected by {\it Swift}-UVOT.

Due to the limited number of data points in the UVOT observations (6 per individual orbital phase), our data fitting is restricted to a simple blackbody model. However, \citet{Brasseur2023} argue that such simplistic models may not adequately fit both the NUV and optical flare data simultaneously, considering the time- and wavelength-dependent optical depths of flare emission in the lower stellar atmosphere. We confirm their assertion by finding that fitting DQ~Tau's combined NUV and optical flare data with a one-temperature blackbody model yields a poor fit (not shown here). Therefore, our focus shifts towards fitting the NUV and optical components of the flares separately.

For the orbital range $\Phi = (0.96 - 1.1)$, a preliminary estimation is conducted to determine the combined energy of the main and secondary flares during periastron. Firstly, the $V$-, $B$-, and $U$-band UVOT fluxes are adjusted by subtracting two different baseline levels, as indicated by the dashed and dotted lines in Figure~\ref{fig:nustar_swift_lcs}c. The start and end points of the optical/NUV flares are determined through visual inspection of the UVOT light curves in Figure~\ref{fig:nustar_swift_lcs}. These points correspond to orbital phase instances with significant changes in the decay slopes of the UVOT light curves. The UVOT data outside the flares were used in the regression analysis to obtain baseline polynomial fits.

\begin{figure*}
\centering
\includegraphics[width=0.9\textwidth]{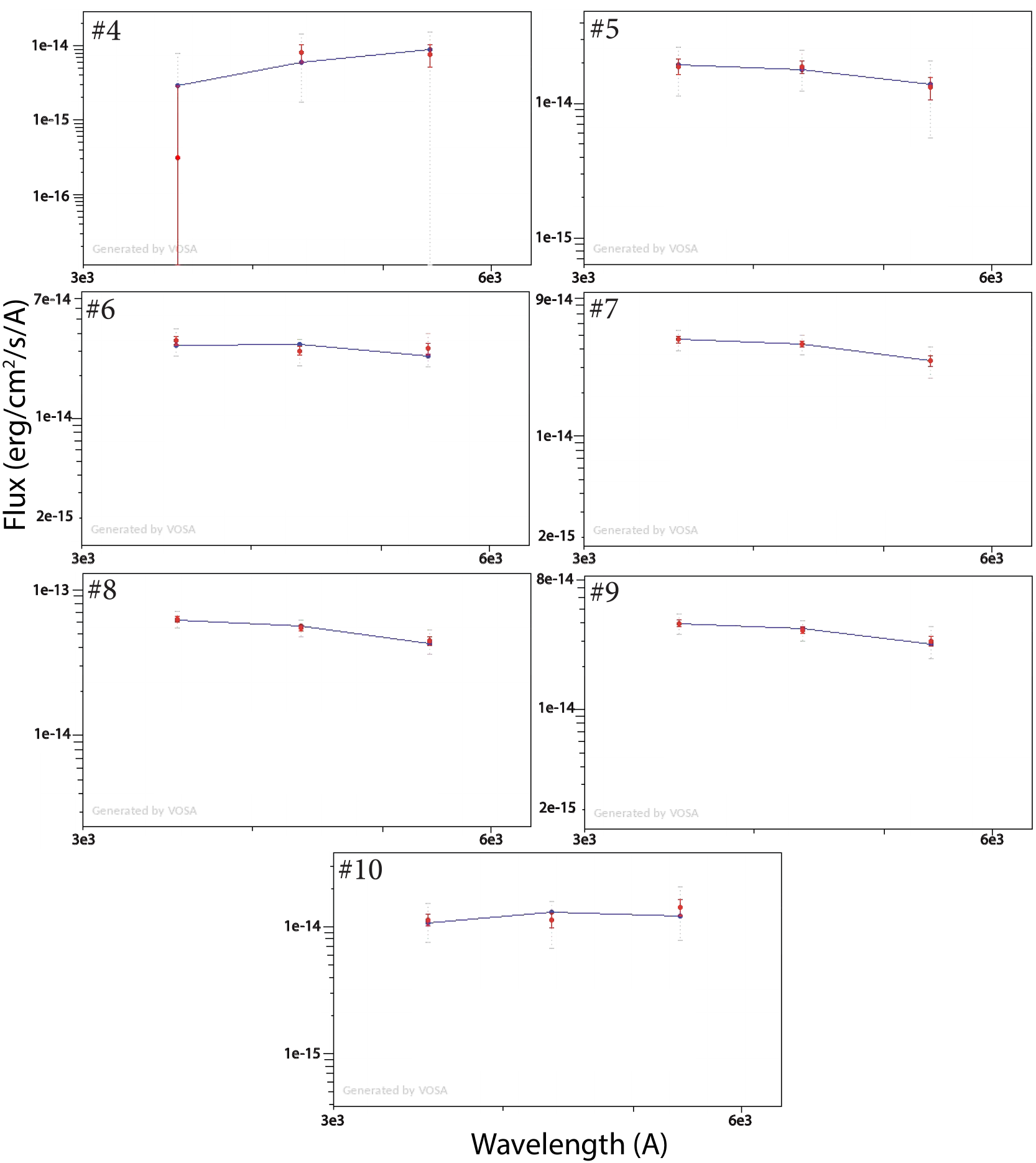}
\caption{Fitting of the optical SEDs using the blackbody model. The fitted data are represented in red, while the 3-$\sigma$ error bars are displayed in grey. The best-fit blackbody models are illustrated in blue. Figure legends show the sequential numbers listed in Column~1 of Table~\ref{tab:optical_seds}.} \label{fig:optical_seds}
\end{figure*}

Subsequently, the VOSA SED (spectral energy distribution) analyzer \citep{Bayo2008} is utilized to fit these two sets of flux data using the blackbody model (Figure~\ref{fig:optical_seds}). The visual source extinction is fixed at $A_V=1.7$~mag \citep{Fiorellino2022}. The effective temperature model parameter is constrained within a range of $(4000-14000)$~K, which aligns with the temperatures observed in Solar and stellar flares \citep{Kowalski2013,Flaccomio2018}, as well as the temperatures found in accretion hot-spots \citep{Tofflemire2017}. The best-fit results for the version with the polynomial baseline are presented in Table~\ref{tab:optical_seds} and Figure~\ref{fig:optical_seds}. For most data points, the inferred temperatures closely align with $T \sim 8000$~K, consistent with the assumed temperatures of $T=9000-10000$~K for the Sun and young NGC~2264 stars \citep{Kretzschmar2011,Flaccomio2018}. For the version with the linear baseline, the inferred temperatures appear even closer to $T \sim 10000$~K. However, it is important to note that such temperature values are also applicable to accretion hot spots on DQ~Tau and other stars \citet{Tofflemire2017}. The formal statistical errors on the inferred bolometric luminosities are less than 1\%.

\begin{figure*}
\centering
\includegraphics[width=0.9\textwidth]{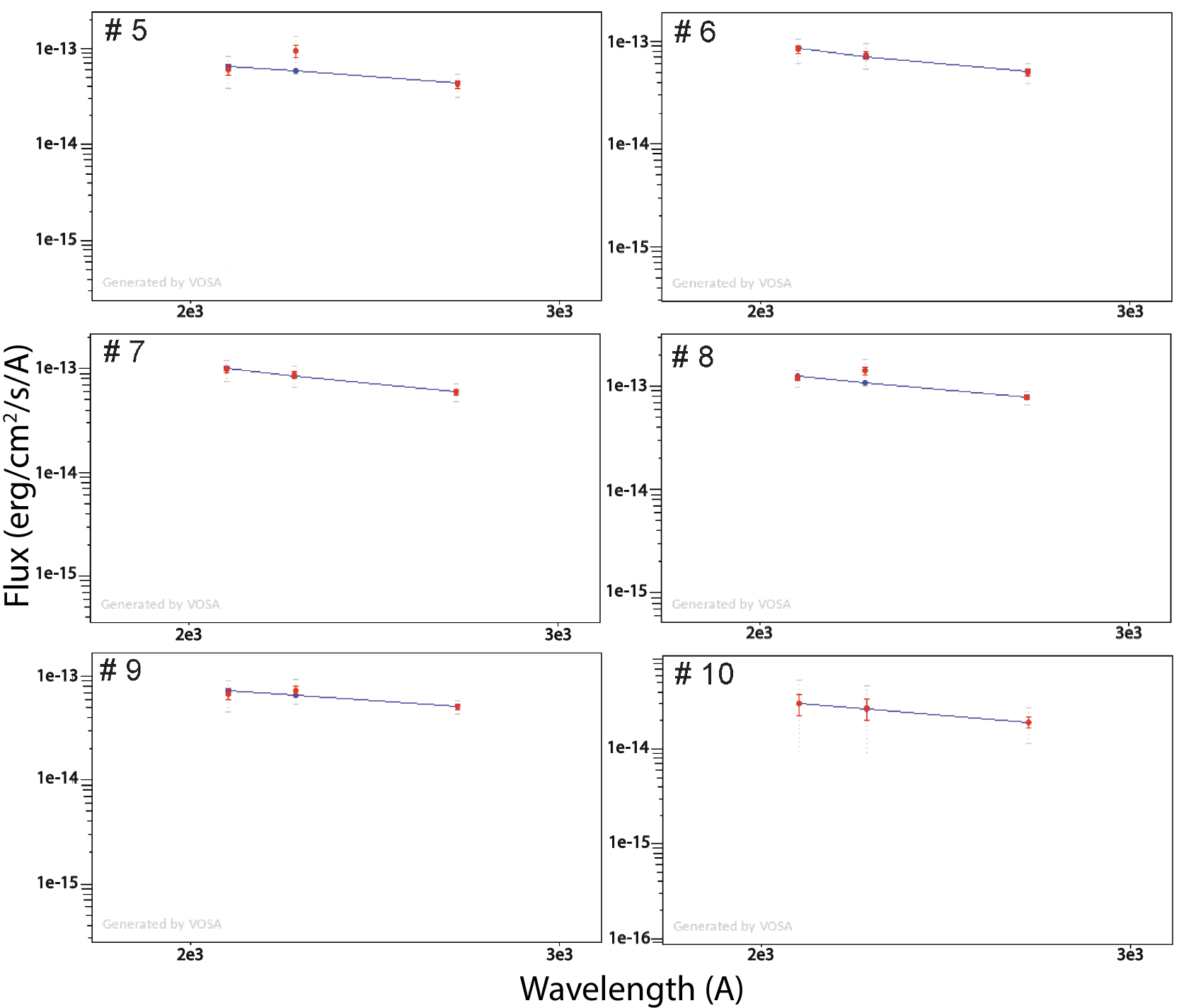}
\caption{Fitting of the NUV SEDs using the blackbody model. The fitted data are represented in red, while the 3-$\sigma$ error bars are displayed in grey. The best-fit blackbody models are illustrated in blue. Figure legends show the sequential numbers listed in Column~1 of Table~\ref{tab:nuv_seds}.} \label{fig:nuv_seds}
\end{figure*}

The fitting procedure for the NUV flare is performed similarly, with the initial blackbody temperature parameter upper boundary raised to $36000$~K as considered for GALEX-NUV stellar flares in \citet{Brasseur2023}. Such a range also includes the temperature value of $25000$~K proposed for the UV component of solar ``white-light'' flares \citep{Fletcher2007}. The best-fit results for the version with the polynomial baseline are presented in Table~\ref{tab:nuv_seds} and Figure~\ref{fig:nuv_seds}. The temperatures inferred for the DQ~Tau flares fall within the range of $(18000-26000)$~K, closely resembling the high blackbody temperatures typically associated with solar flares. However, it is worth noting that these elevated temperatures may also be attributed to the accreting material in the vicinity of the shock regions \citep{SiciliaAguilar2015}.  

Next, the total energies emitted by both periastron flares in the optical ($380-700$~nm; $E_{opt}$) and NUV ($177.1-283.1$~nm assuming GALEX band; $E_{NUV}$) are estimated by integrating the bolometric luminosity values obtained from the fits over the duration of the flares. The bolometric flare energies are determined to lie within the range $E_{bol} = (1-2) \times 10^{38}$~erg. These resulting energies are then reduced by 60\% and 70\% to account for contributions outside the optical and NUV wavelength ranges of the associated blackbody spectra, respectively.

In the case of the polynomial baseline, the total estimated optical energy emitted during the periastron flares within the orbital phase $\Phi=(0.96-1.1)$ reaches $E_{opt} = 7 \times 10^{37}$~erg. Conversely, for the linear baseline version, $E_{opt}$ amounts to $5 \times 10^{37}$~erg. As for the NUV energies, the inferred values for these flares are $E_{NUV} = 6 \times 10^{37}$~erg and $E_{NUV} = 4 \times 10^{37}$~erg for the polynomial and linear baseline versions, respectively.

Although the formal statistical errors associated with the optical and NUV flare emission energy values are relatively small (less than 1\%), it is crucial to acknowledge the presence of several systematic factors that contribute to the uncertainty of these energies. These factors stem from various sources, including the limitations imposed by the cadence of the {\it Swift}-UVOT observations, the potential influence of high accretion rates during magnetic reconnection events \citep{Tofflemire2017}, the selection of baseline levels, and the necessity of making certain assumptions in the emission model. These assumptions encompass considerations such as distinguishing between emission lines and continuum, as well as determining whether the emission is optically thin or optically thick \citep{Kowalski2013, Flaccomio2018, Brasseur2023}.

Only a few instances of simultaneously observed NUV and optical stellar flares have been documented in the existing literature \citep[see Table 7 in][]{Brasseur2023}. Notably, certain M-type stars such as GJ~1243, AD~Leo, and DG~CVn have been identified as exhibiting comparable NUV and optical flare energies. Similarly, in our investigation, we uncover comparable NUV and optical energies for the significant periastron flares in DQ~Tau, with $E_{opt} = (5-7) \times 10^{37}$~erg and $E_{NUV} = (4-6) \times 10^{37}$~erg. However, it is important to note that the DQ~Tau flares surpass the energy levels of the aforementioned M-type star flares by $2-7$ orders of magnitude.

The optical-to-X-ray energy ratio for the periastron flares of DQ~Tau falls within the range of $E_{opt}/E_{X} = (75-370)$. This ratio surpasses the energy ratios calculated for a few powerful solar flares by \citet{Woods2006}, which span a range of $E_{opt}/E_{X,(0.5-8)keV} = (25-40)$, when converting the GOES flare energies reported in Woods et al. to X-ray energies in the $(0.5-8)$~keV band \citep{Flaccomio2018}. Furthermore, DQ~Tau's $E_{opt}/E_{X}$ energy ratio notably exceeds the ratio of $E_{opt}/E_{X} \sim 10$ observed in a few optical-X-ray CoRoT-{\it Chandra} flares with $E_{X} \ge 10^{35}$~erg generated by young stellar members of the NGC~2264 star forming region \citep{Flaccomio2018}.

\begin{deluxetable*}{cccccccc}
\tabletypesize{\normalsize}
\tablecaption{Fitting Optical SEDs \label{tab:optical_seds}}
\tablewidth{0pt}
\tablehead{
\colhead{\#} & \colhead{$\Phi$} & \colhead{$F_{\nu,V}$} & \colhead{$F_{\nu,B}$} & \colhead{$F_{\nu,U}$} &  \colhead{$\chi^{2}$} & \colhead{$T$} & \colhead{$L_{bol}$} \\
\colhead{} & \colhead{} & \colhead{(mJy)} & \colhead{(mJy)} &  \colhead{(mJy)} & \colhead{} & \colhead{(K)} & \colhead{($L_{\odot}$)}\\
\colhead{(1)} & \colhead{(2)} & \colhead{(3)} & \colhead{(4)} & \colhead{(5)} & \colhead{(6)} & \colhead{(7)} & \colhead{(8)}
}
\startdata
4  & 0.96  & $1.53\pm0.51$ & $0.65\pm0.17$ & $0.01\pm0.08$ & 2.13 & 4400 & 0.12\\
5  & 0.98  & $2.60\pm0.50$ & $1.51\pm0.17$ & $0.60\pm0.08$ & 0.25 & 8000 & 0.13\\
6  & 0.99  & $6.19\pm0.54$ & $2.41\pm0.17$ & $1.12\pm0.08$ & 5.36 & 7400 & 0.25\\
7  & 1.02  & $6.60\pm0.53$ & $3.49\pm0.18$ & $1.48\pm0.08$ & 0.07 & 8050 & 0.31\\
8  & 1.04  & $8.78\pm0.57$ & $4.37\pm0.19$ & $1.99\pm0.09$ & 1.07 & 8200 & 0.40\\
9  & 1.06  & $5.92\pm0.49$ & $2.86\pm0.16$ & $1.26\pm0.07$ & 0.71 & 7950 & 0.26\\
10  & 1.09  & $2.80\pm0.42$ & $0.90\pm0.12$ & $0.36\pm0.04$ & 2.44 & 6350 & 0.11\\
\enddata 
\tablecomments{Column 1: Sequential number of the data points shown in Figure~\ref{fig:nustar_swift_lcs}c and assigned in Figure~\ref{fig:optical_seds}. Column 2: Orbital phase. Columns 3-5: UVOT fluxes in the $V$-, $B$-, and $U$-bands. Column 6: $\chi^2$ representing the goodness-of-fit of the data to the blackbody model. Columns 7-8: Effective temperature and bolometric luminosity inferred from the fitting process.}
\end{deluxetable*}

\begin{deluxetable*}{cccccccc}
\tabletypesize{\normalsize}
\tablecaption{Fitting NUV SEDs \label{tab:nuv_seds}}
\tablewidth{0pt}
\tablehead{
\colhead{\#} & \colhead{$\Phi$} & \colhead{$F_{\nu,W1}$} & \colhead{$F_{\nu,M2}$} & \colhead{$F_{\nu,W2}$} &  \colhead{$\chi^{2}$} & \colhead{$T$} & \colhead{$L_{bol}$} \\
\colhead{} & \colhead{} & \colhead{(mJy)} & \colhead{(mJy)} &  \colhead{(mJy)} & \colhead{} & \colhead{(K)} & \colhead{($L_{\odot}$)}\\
\colhead{(1)} & \colhead{(2)} & \colhead{(3)} & \colhead{(4)} & \colhead{(5)} & \colhead{(6)} & \colhead{(7)} & \colhead{(8)}
}
\startdata
5  & 0.98  & $0.33\pm0.03$ & $0.14\pm0.02$ & $0.08\pm0.01$ & 7.69 & 20000 & 0.24\\
6  & 0.99  & $0.39\pm0.03$ & $0.11\pm0.01$ & $0.11\pm0.01$ & 0.28 & 26000 & 0.36\\
7  & 1.02  & $0.47\pm0.03$ & $0.13\pm0.01$ & $0.13\pm0.01$ & 0.26 & 25000 & 0.41\\
8  & 1.04  & $0.61\pm0.03$ & $0.21\pm0.02$ & $0.16\pm0.01$ & 6.39 & 23000 & 0.49\\
9  & 1.06  & $0.40\pm0.02$ & $0.11\pm0.01$ & $0.09\pm0.01$ & 1.54 & 18650 & 0.25\\
10  & 1.09  & $0.15\pm0.02$ & $0.04\pm0.01$ & $0.04\pm0.01$ & 0.01 & 23000 & 0.12\\
\enddata 
\tablecomments{Column 1: Sequential number of the data points shown in Figure~\ref{fig:nustar_swift_lcs}d and assigned in Figure~\ref{fig:nuv_seds}. Column 2: Orbital phase. Columns 3-5: UVOT fluxes in the $UVW_1$-, $UVM_2$-, and $UVW_2$-bands. Column 6: $\chi^2$ representing the goodness-of-fit of the data to the blackbody model. Columns 7-8: Effective temperature and bolometric luminosity inferred from the fitting process.}
\end{deluxetable*}
\clearpage

\subsubsection{Possible X-ray Energy Range for The {\it Chandra} Flare} \label{sec:chandra_flare_energy}

Given the significant time gaps of approximately 1 day between individual {\it Chandra} observations, the morphology and time-scales of the {\it Chandra} flare remain poorly constrained. However, the observed peak of the flare, adjusted for the baseline, reaching $5 \times 10^{31}$~erg~s$^{-1}$, strongly suggests its classification as a super-flare, assuming its nature is similar to the numerous X-ray super-flares studied in \citet{Getman08a,Getman2021}. By considering the second highest count rate data point from {\it Chandra} as part of the same super-flare, an upper limit for the flare's decay time-scale is estimated to be approximately $\tau_{decay} \sim 40$~ksec (see \S\ref{sec:xray_lcs}). Conversely, if the flare's energy is situated at the lower end of the previously studied X-ray super-flares, the lower limit on $\tau_{decay}$ would be around 10~ksec \citep{Getman08a,Getman2021}. Consequently, the X-ray energy of the {\it Chandra} flare is anticipated to fall within the range of $E_X \sim L_{X,pk} \times \tau_{decay} \sim (5-20) \times 10^{35}$~erg. Nevertheless, due to the limited amount of {\it Chandra} data, even this wide energy range should be considered with caution.

\subsubsection{Flaring Coronal Loop Length Near Periastron} \label{sec:loop_length}

Within the framework of the time-dependent hydrodynamic model proposed by \citet{Reale1997, Reale2007}, a preliminary estimation of the coronal loop length can be made for the primary X-ray periastron flare. This model is applicable in cases where flaring multi-loop arcades are present, with the flare being predominantly governed by a single loop or multiple individual loop events that exhibit similar temperature and emission measure temporal profiles, occurring almost simultaneously \citep{Reale2014, Getman2011}.

The hydrodynamic simulations presented by \citet{Reale1997} establish a relationship between loop height and three observable parameters during the flare decay phase: the exponential timescale for flare decay, $\tau_{decay}$; the plasma temperature at the peak emission measure, $T_{EM,pk}$; and the slope on the log-temperature versus log-density diagram, denoted as $\zeta$. The half-length of a coronal loop, $L_{decay}$, can be expressed as $L_{decay} \propto \tau_{decay} T_{EM,pk}^{1/2} /F(\zeta)$, where $F(\zeta)$ is a function that accounts for prolonged heating. Higher values of $\zeta$ correspond to freely decaying loops without sustained heating, while lower values indicate loops with prolonged heating.

Due to the relatively limited counting statistics of our X-ray data from {\it Swift}-XRT and {\it NuSTAR}, conducting detailed time-resolved spectroscopy is not feasible. However, it is noteworthy that the time-averaged {\it Swift+NuSTAR} solution for the hot plasma component, with $kT_{2} \sim 3$~keV as inferred from our two-temperature model fit (see Table~\ref{tab:xray_spectral_results}), aligns with the time-integrated hottest components of $kT_2 \sim 1.9$~keV and $kT_3 \sim 4.3$~keV observed during the large X-ray {\it Chandra} flare that occurred during the 2010 periastron passage \citep{Getman2011}. The X-ray energies and $\tau_{decay}$ timescales for these two flares also exhibit close similarities\footnote{The X-ray energy of the 2010 X-ray flare, as reported by \citet{Getman2011}, has been adjusted to account for the updated {\it Gaia} distance of 195~pc. This correction results in an intrinsic peak X-ray luminosity of $L_{X,pk} = 7 \times 10^{30}$~erg~s$^{-1}$. Taking into consideration the flare's decay time-scales of $\tau_{decay} \sim 41$~ksec \citep{Getman2011}, the total flare energy can be estimated as $L_{X,pk} \times \tau_{decay} \sim 3 \times 10^{35}$~erg. Using a refined calculation method, integrating X-ray luminosity ($L_X$) over the flare duration and accounting for truncation by {\it Chandra} exposure, we estimate the flare energy more accurately as $E_X = (3.1 \pm 0.2) \times 10^{35}$~erg.}, with $E_X \sim 3 \times 10^{35}$~erg and $\tau_{decay} \sim 40-50$~ksec. Considering reasonable assumptions of comparable peak flare plasma temperatures and amount of sustained heating, it further suggests that the coronal structure associated with the main flare near the July 30, 2022 periastron passage may possess a height similar to that observed during the 2010 epoch, spanning a few to several stellar radii.

\begin{figure*}
\centering
\includegraphics[width=0.8\textwidth]{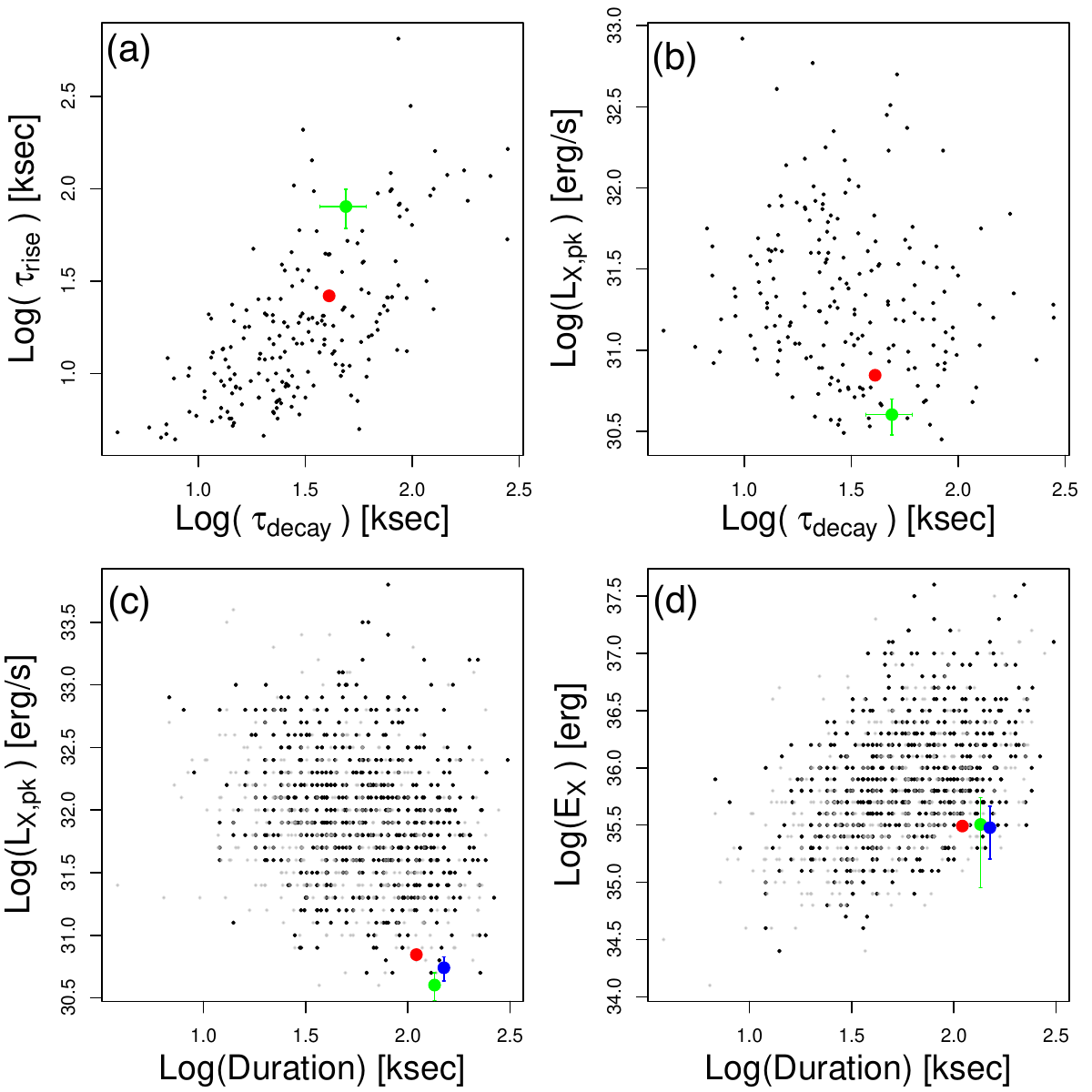}
\caption{Comparison of the periastron-flare properties of DQTau (colored points) with those of large X-ray flares from young stellar members of the Orion Nebula Cluster, known as COUP flares \citep{Getman08a}, and of numerous other star-forming regions, referred to as MYStIX/SFiNCs flares \citep{Getman2021}. The DQ~Tau X-ray flares are color-coded as follows: the 2010 flare \citep[][red]{Getman2011}, the 2021 flare \citep[][blue]{Getman2022b}, and the 2022 flare (current study, green). Panels (a) and (b) display the flare rise/decay time-scales and peak X-ray luminosities for COUP flares (black) and DQ~Tau flares (colored points). Panels (c) and (d) show the flare duration, peak X-ray luminosity, and flare energy for MYStIX/SFiNCs flares (black and grey) and DQ~Tau flares (colored points). The lower limits for the ``incomplete'' (partially captured) MYStIX/SFiNCs flares are shown in grey.} \label{fig:comparison_with_coup_mystix_sfincs}
\end{figure*}

\section{Comparison with X-ray Flares from Young Stars} \label{sec:comparison_with_coup_mystix_sfincs}

We conducted three observations of the periastron passages of DQ~Tau using X-ray telescopes: in 2010 \citep{Getman2011}, 2021 \citep{Getman2022b}, and 2022 (current study). Notably, significant X-ray flares were detected on all three occasions. 

In Figure~\ref{fig:comparison_with_coup_mystix_sfincs}, we compare the duration, peak X-ray luminosity, and energetics of these DQ~Tau flares with numerous large X-ray flares produced by young stars, as studied by \citet[][also known as COUP flares]{Getman08a} and \citet[][also known as MYStIX/SFiNCs flares]{Getman2021}. In Figure~\ref{fig:comparison_with_coup_mystix_sfincs}, we only include the ``main'' DQ~Tau X-ray flares that occur roughly within the orbital phase range of $(0.95-1.05)$. However, it is important to note that more X-ray flaring events of comparable energetics are present within the $(1.05-1.2)$ orbital phase range (see Figure~\ref{fig:nustar_swift_lcs} here and Figure~1 in \citet{Getman2022b}).

It is also noteworthy that somewhat different methodologies for flare detection and analysis were employed in \citet{Getman08a} and \citet{Getman2021}, due to distinct scientific objectives pursued in these two papers. 

As a result, \citet{Getman08a} reported the rise and decay timescales, flare peak X-ray luminosities, but no energies for the COUP flares, while \citet{Getman2021} reported only total flare durations (with no differentiation between rise and decay), flare peak X-ray luminosities, and energies for the MYStIX/SFiNCs flares.

Consequently, the rise and decay timescales, as well as flare peak X-ray luminosities, are compared between the DQ~Tau and COUP flares (see Figure \ref{fig:comparison_with_coup_mystix_sfincs}a and b), while the flare durations, peak luminosities, and energies are compared between the DQ~Tau and MYStIX/SFiNCs flares (see Figure \ref{fig:comparison_with_coup_mystix_sfincs}c and d).

Moreover, the limited availability of individual observations from the {\it Swift}-XRT associated with the DQ~Tau 2021 periastron flare (represented by the blue point in the figure) prevented the determination of rise/decay time-scales in \citet{Getman2022b}. As a result, the two upper figure panels do not include the blue point. The figure illustrates DQ~Tau's $L_{X,pk}$ and $E_{X}$ values, which have been adjusted to account for the baseline. To estimate DQ~Tau's flare durations, the time difference between the decay and rise flare tail points with the lowest count rates was considered.

Figure~\ref{fig:comparison_with_coup_mystix_sfincs} illustrates that the DQ~Tau flares lie within the loci of the COUP and MYStIX/SFiNCs flares, albeit having relatively long durations and relatively low peak flare X-ray luminosities. This places them at the sensitivity limit border of the COUP and MYStIX/SFiNCs flare surveys.

\section{Discussion} \label{sec:discussion}

\subsection{X-ray Flares Produced by DQ Tau} \label{sec:discussion_flares}

\subsubsection{Flaring Near Periastron} \label{sec:xray_flares_near_periastron_discussion}

Remarkably, all three main DQ~Tau X-ray flares detected in 2010, 2021, and 2022  near an orbital phase of 1 exhibit similar X-ray energies ($E_X \sim 3 \times 10^{35}$~erg). This energy value is typical of numerous large X-ray flares from young stars (Figure~\ref{fig:comparison_with_coup_mystix_sfincs}). This finding suggests the presence of a consistent powerful source of energy that fuels such flares at each periastron passage. This finding supports the notion previously proposed by \citet{Salter2010,Getman2011} that the magnetosphere collision mechanism is the primary source of magnetic energy powering DQ~Tau's periastron flares.

Our discovery of three super-flare events with energies of $E_X \sim 3 \times 10^{35}$~erg during three distinct periastron passages of DQ~Tau is in direct contradiction with the frequency of such powerful X-ray flares observed in individual MYStIX/SFiNCs stars. Specifically, the anticipated occurrence rate for flares with $E_X \sim 3 \times 10^{35}$~erg in single stars of $\le 1$~M$_{\odot}$ is approximately $3-8$ flares per year per star \citep[][see their equation (4)]{Getman2021}. This rate is significantly lower than our observation of a flare rate of approximately three flares per five days of X-ray observations per star during DQ~Tau's periastrons. This stark contrast provides independent support for the magnetosphere collision mechanism as the primary source of DQ~Tau's periastron flares.

Conceptually, flare-related events triggered by magnetosphere collision may proceed in a manner described by the classical non-thermal thick-target model, potentially involving larger-scale coronal structures. According to the classical non-thermal thick-target model, which applies to solar and stellar flares \citep{Brown1971, Lin1976}, electrons are accelerated to high energies through coronal magnetic reconnection processes. These energetic electrons spiral along the coronal magnetic field lines, emitting radio and microwave radiation (detected in DQ~Tau by \citet{Salter2008,Salter2010}), and subsequently collide with the underlying atmosphere. These collisions result in the production of non-thermal hard X-rays, which may be detectable in the \textit{NuSTAR} energy band. Furthermore, this electron-atmosphere interaction leads to heating of the surrounding transition region, chromosphere, and photosphere plasma, giving rise to the production of optical/ultraviolet (observed by {\it Swift}-UVOT), and infrared radiation. Additionally, the interaction drives chromospheric evaporation, causing the filling of coronal loop(s) with hot plasma that emits thermal X-rays in the soft bands observed by \textit{Chandra/Swift}-XRT.

The Neupert effect, which establishes a correlation between the time-integrated radio or microwave (or hard non-thermal X-ray) light curve and the rising portion of the soft X-ray light curve \citep{Neupert1968}, serves as compelling observational evidence supporting the classical non-thermal thick-target model. This effect has been observed in numerous solar flares \citep[e.g.,][]{Dennis1993} and certain stellar flares \citep[][]{Gudel2002}. Remarkably, \citet{Getman2011} discovered the presence of the Neupert effect in the context of the January 11-12, 2010 periastron passage of DQ Tau, where they observed correlations between the IRAM microwave and {\it Chandra} X-ray flares. Furthermore, \citet{Salter2010} and \citet{Getman2011} found that the heights of the coronal structures associated with these flares reached several stellar radii.

Based on the analysis of nearly simultaneous mm-band and X-ray flares observed during the 2010 periastron passage of DQ~Tau, and using the framework of the generalized Neupert effect proposed in \citet{Guedel1996}, \citet{Getman2011} derived an estimation of the kinetic energy rate injected into the chromosphere by non-thermal electrons. Taking into account the updated {\it Gaia} distance, the estimated value falls in the range of $(1-2) \times 10^{32}$~erg~s$^{-1}$. If such injection rate persists throughout the $(0.95-1.1)$ orbital phase range of DQ~Tau, it can generate radiation with a total energy of $(2-4) \times 10^{37}$~erg.
 
The analytic model proposed by \citet{Adams2011} provides a comprehensive description of the magnetic energy release process in eccentric binary systems, specifically addressing the stored magnetic energy within the large-scale, dipole magnetic fields of the stellar components. This release is achieved through the magnetic interaction of the binary stellar components' magnetospheres. Additionally, the authors discuss the replenishment of this magnetic energy through the combined effects of the orbital and spin motions of the binary components. \citet{Adams2011} and \citet{Das2023} have determined that this magnetic model yields reasonable estimates of the magnetic reconnection energy responsible for powering the radio and X-ray flares observed near the periastrons of the V773~Tau and $\epsilon$~Lupi eccentric binaries, respectively.

By employing the magnetic model introduced by \citet{Adams2011}, our analysis unveils a substantial amount of magnetic energy that explains the observed flaring phenomena across multiple bands, including mm, optical, NUV, and X-ray. In our analysis, we consider the following orbital and stellar parameters for DQ~Tau: an orbital period of $P_{orb}=15.8$~days, a semimajor axis of $a_0=0.142$~au, an eccentricity of $\epsilon = 0.58$, a stellar radius of $R_{\star}=2$~R$_{\odot}$, and a surface magnetic field strength of $B_{\star} = 2.5$~kG \citep{Salter2010, Czekala2016, Fiorellino2022,Pouilly2023}.

\begin{figure}
\centering
\includegraphics[width=0.5\textwidth]{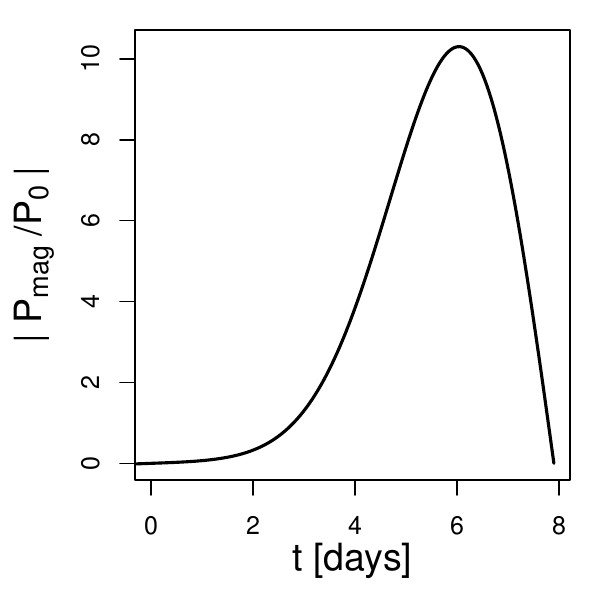}
\caption{$|P_{mag}/P_{0}|$ as a function of days since DQ Tau's apastron, over one-half orbit. Refer to Figure~5 in \citet{Adams2011} for a similar analysis in the case of the V773~Tau~A binary system.} \label{fig:pmax_to_p0}
\end{figure}

Equation (17) in \citet{Adams2011} provides an estimate of the magnetic energy release rate ($P_{mag}$) resulting from magnetosphere interaction. $P_{mag}$ is a function of two components: the fiducial scale $P_0 = (2 \pi B_{\star}^2 R_{\star}^6) / 
(P_{orb} a_0^3) = 2.2 \times 10^{31}$~erg~s$^{-1}$ and the function of eccentricity and orbital angle, $f(\epsilon, \Theta)$. Notice that the eccentricity of DQ~Tau is twice as high as that of V773~Tau~A, which causes the function $f(\epsilon, \Theta)$ to reach a value of 10 in the case of DQ~Tau (Figure~\ref{fig:pmax_to_p0}), but to remain below 0.6 in the case of V773~Tau~A (Figure~5 in \citet{Adams2011}). Within the 2.4-day, 3.5-day, and 4.0-day windows of maximum energy release\footnote{The 2.4-day window encompasses the first two X-ray flares and the optical/NUV flare, while the 3.5-day and 4.0-day windows additionally include the third and fourth X-ray flare.}, the average values of $f(\epsilon, \Theta)$ are 9.1, 8.0, and 7.2  respectively. 

Within the context of the basic equation (17) and Figure 5 in Adams et al. (2011), as well as Figure \ref{fig:pmax_to_p0} presented here, it's worth noting that during the orbital phase range of $(0.5-1)$, the two stars draw closer to each other. This proximity leads to the compression of their magnetic fields, resulting in an excess of magnetic energy becoming available to fuel flare events. Conversely, it's not expected for there to be a significant release of magnetic energy during the orbital phase range of $(0-0.5)$ when the stars move away from each other. During this phase, it is anticipated that the two magnetospheres will replenish their energies through a combination of orbital and stellar spin motions, as well as through internal stellar dynamos.

According to this simplified model, the peak of the available excess magnetic energy occurs 2 days before reaching the periastron point (as shown in Figure~\ref{fig:pmax_to_p0}). Equation (17) and Figure 5 in Adams et al. (2011) assume that the magnetic configuration can instantaneously adjust to magnetic stresses and immediately dissipate excess energy. However, in reality, magnetic accumulation and reconnection do not occur instantaneously. Therefore, related flare events may be observed near and after the periastron point (Fred Adams, private communication).

Using this equation, we calculate a magnetic energy release rate of $\sim 2 \times 10^{32}$ erg s$^{-1}$ during any of the 2.4-day, 3.5-day, and 4.0-day periods of maximum energy release near periastron. This corresponds to a total magnetic energy release near periastron of approximately $E_{mag} = 4.1 \times 10^{37}$~erg, $5.3 \times 10^{37}$~erg, and $5.5 \times 10^{37}$~erg within the 2.4-day, 3.5-day, and 4.0-day windows, respectively.

These model-predicted magnetic energy release rate and the total released magnetic energy within the orbital phase range of $0.95-1.1$, i.e. 2.4-day window, are found to be sufficient to sustain the energy rate injection by non-thermal electrons, estimated to be $(1-2) \times 10^{32}$~erg~s$^{-1}$ (as derived from our mm-band and X-ray band flares observed in 2010). Moreover, the model accounts for the X-ray flaring with an energy of $E_X \sim 6 \times 10^{35}$~erg. This includes the combined main, second, and third flares within the 3.5-day window (\S\ref{sec:flare_energy_xray}). If the energy of the fourth X-ray flare is on par with that of the initial three flares, the model will readily account for the fourth event as well.

However, a noticeable discrepancy arises between the inferred optical and NUV (as well as the corresponding bolometric) flare energies and the energies associated with magnetic reconnection and non-thermal electrons. The fitting of the optical/NUV flares in \S~\ref{sec:flare_energy_optical} yields a bolometric energy for the flares across the $(0.95-1.1)$ orbital phase of approximately $E_{bol} \sim (1-2) \times 10^{38}$~erg, which is a factor of $(2-5)$ higher than the predicted total magnetic reconnection energy of $E_{mag} \sim 4 \times 10^{37}$erg. 

While it is plausible that the presence of general uncertainties surrounding all flare energetics and stellar properties, as derived in the current study or obtained from the literature, might account for part of this discrepancy, we propose two main sources of this inconsistency. The first is related to the partial inclusion of accretion-related optical/NUV emission \citep{Tofflemire2017, Kospal2018, Muzerolle2019, Fiorellino2022}. The second is associated with the potential overestimation of temperatures and bolometric luminosities in our simplistic modeling of the optical/NUV periastron flaring (see \S~\ref{sec:flare_energy_optical}). Specifically, previous research has reported the presence of optically thin line and Balmer continuum emissions in the $U$-band radiation of stellar flares \citep[Figure~3 in][]{Kowalski2013}. Since our model fitting does not account for such ``additional'' emission, it may lead to an overestimation of the temperature and luminosity of the true black body component. 

The temperatures around $T \sim 10000$~K, which are obtained from our optical SED analyses, are relevant in both cases: stellar magnetic reconnection flares and accretion hot spots. Unfortunately, neither temperatures nor colors\footnote{The $U,B,V$ UVOT colors (and magnitudes) across the entire $(0.9-1.25)$ orbital phase range covered by the {\it Swift} observations are consistent with Figure~1 of \citet{Tofflemire2017} and indicate that DQ~Tau is bluer when brighter and redder when fainter.} (used as observational proxies for temperatures) can distinguish between flare and accretion events. The morphology of large optical flares, as observed in young stars of NGC~2264 \citep[][their Appendix~B]{Flaccomio2018}, often differs from the 'fast-rise and slow-decay' morphology, and their durations frequently align with those of concurrent soft X-ray flares. Similar optical-X-ray behavior is observed for both the main and second DQ~Tau flares (see Figure~\ref{fig:nustar_swift_lcs}). The high durations of the optical, NUV, and X-ray flares during periastron in DQ~Tau can be explained by the sustained heating resulting from the magnetic energy release due to colliding magnetospheres. The presence of two distinct optical/NUV peaks, particularly evident in the $V$, $W_1$, and $M_2$ bands (Figure~\ref{fig:nustar_swift_lcs}), occurring before the peaks of the primary and secondary X-ray flares, is indicative of an observational feature associated with the solar/stellar flare Neupert effect.  Overall, we identify observational indications suggesting the presence of magnetic reconnection-related optical/NUV flares, but it is not possible to differentiate their energetics from those of the underlying accretion events based solely on the optical/NUV data.

In the comparison of observational optical-to-X-ray energy ratios, it is found that the ratio for the DQ~Tau flares is significantly higher than those for large solar flares and flares from young NGC~2264 stars (see \S \ref{sec:flare_energy_optical}). However, when reducing the values of $E_{bol}$, $E_{opt}$, and $E_{NUV}$ by a factor of 5, they align with the magnetic reconnection energy predicted by the \citet{Adams2011} model. Simultaneously, this adjustment lowers the optical-to-X-ray energy ratio for the periastron flares of DQ~Tau to $E_{opt}/E_{X} = (15-75)$, bringing it in line with ratios observed in solar flares and more energetic flares from young stars \citep{Woods2006, Flaccomio2018}. 
 

In summary, distinguishing between the energetics of magnetic reconnection-related and accretion-related periastron events based on the optical/NUV data alone is challenging. Nevertheless, we have observed the Neupert effect between millimeter and X-ray flares, and identified consistent rates of magnetic energy release and non-thermal electron injection. The optical-to-X-ray energy ratios of DQ Tau and large solar/stellar flares align when we adjust the observed optical energy of DQ Tau to match the energy levels predicted by the \citet{Adams2011} model. Furthermore, two distinct optical/NUV peaks precede the corresponding X-ray peaks, and the optical and X-ray events powered by sustained heating from colliding magnetospheres have similar durations. These findings collectively support the idea that the millimeter/X-ray periastron flares, and tentatively, the magnetic reconnection-related components of the optical/NUV emissions, conform to the classical solar/stellar non-thermal thick-target model.

\subsubsection{Non-Periastron Flaring} \label{sec:xray_flares_away_from_periastron_discussion}

In this section, we discuss speculatively the possible origin of the X-ray flaring events observed outside of DQ~Tau's periastron passage. We have gathered observational evidence of two distinct non-periastron X-ray flares thus far: one from the {\it Swift}-XRT data acquired in 2017 (refer to Figure~1 in \citealt{Getman2022b}), and another from our {\it Chandra} observation conducted in 2022 (Figure~\ref{fig:chandra_lc} in this paper). If the sparse {\it Swift}-XRT data points depicted in Figure~1 of \citet{Getman2022b} are associated with the same X-ray flare event, the estimated flare energy could reach up to $E_X \sim 9 \times 10^{35}$~erg. Regarding the 2022 {\it Chandra} flare, the expected energy range falls within $E_X \sim (5-20) \times 10^{35}$~erg (refer to \S~\ref{sec:chandra_flare_energy}). Such large X-ray flares in single stars exhibit an occurrence rate of approximately $(0.2-5)$ flares per year per star, significantly lower than the rate of 2 flares per 18-day span of X-ray observations for DQ~Tau. Therefore, it is improbable for such energetic flares to occur randomly, suggesting that DQ~Tau's stellar binarity may play a contributing role. 

According to the magnetic model proposed by \citet{Adams2011}, it is plausible that the release of magnetic energy from the large-scale magnetic fields can occur at orbital phases away from periastron, albeit with a lower energy release rate. Moreover, the interaction between the magnetospheres of the two stars could disrupt the small-scale surface magnetic fields, potentially leading to additional flaring events. However, the occurrence of an extended magnetically calm phase detected by {\it Chandra} lasting over 7 days, devoid of significant flares (corresponding to the orbital phase range of $(1.1-1.55)$ in Figure~\ref{fig:chandra_lc}), suggests a more intricate nature underlying non-periastron X-ray flaring.

Numerous less energetic optical flares ($E_{opt} \sim (10^{32}-10^{35})$~keV), believed to be triggered by magnetic reconnection, were also identified during multiple orbits of DQ~Tau by \citet{Kospal2018}. These flares occur independently of the orbital phase of the system.

Given the predictable occurrence of X-ray super-flares and accretion outbursts in close proximity to periastron passage, DQ~Tau stands as an exceptional laboratory for examining the impact of stellar radiation on the gas-phase ion chemistry within its disk. Nevertheless, the system also displays sporadic and frequent super-flaring events away from periastron, thereby rendering a comprehensive multi-wavelength investigation into the influence of DQ~Tau's radiation on its disk a more formidable undertaking than initially envisioned.

\subsection{Non-detection of Hard Non-thermal X-rays} \label{sec:discussion_non_thermal_xrays}
In this study, we performed \textit{NuSTAR} observations in the vicinity of DQ~Tau's periastron in order to investigate the presence of the non-thermal flaring X-ray component, as predicted by the classical non-thermal thick-target model. Our {\it NuSTAR} observation did not reveal any significant hard X-ray emission ($>10$~keV) from DQ Tau near periastron (\S\S~\ref{sec:nustar_data}, \ref{sec:xray_spectra}).

To the best of our knowledge, only a few flares from young stellar objects, all within the nearby $\rho$~Oph region, have been observed by \textit{NuSTAR} thus far \citep{Vievering2019,Pillitteri2019}. Vievering et al. detected several bright flares from IRS43, WL19, and Elias~29 young stellar objects, but no evidence of non-thermal X-ray emission was found. In the case of another two detected X-ray flares from the Elias~29 object, Pillitteri et al. reported a tentative power-law excess of hard X-ray emission in the $(20-50)$~keV band, as deduced from its \textit{NuSTAR} spectrum.

\citet{Isola2007} conducted an analysis of soft X-ray GOES and hard X-ray RHESSI data for 45 bright Solar flares, revealing a strong correlation between the GOES fluxes in the $(1.6-12.4)$~keV band and RHESSI fluxes in two bands, $(20-40)$~keV and $(60-80)$~keV. These findings align with the expectations derived from the thick-target model. Isola et al. further demonstrated that the same scaling law observed for solar flares between the $(1.6-12.4)$~keV and $(20-40)$~keV fluxes also holds true for more powerful stellar flares.

Assuming the scaling law's applicability to the main DQ~Tau flare, we convert the flare's peak X-ray luminosity (corrected for the baseline level of X-ray emission) of $L_{X,pk} = 4 \times 10^{30}$~erg~s$^{-1}$ in the $(0.5-8)$~keV band (Table~\ref{tab:xray_spectral_results}) to a GOES-like flux in the $(1.6-12.4)$~keV band at a distance of 1~au from the system, resulting in $F_{G} = 0.61$~W~m$^{-2}$. According to equation (1) from \citet{Isola2007}, this $F_{G}$ flux predicts an X-ray flux in the $(20-40)$~keV band for the main DQ~Tau flare of approximately $F_{X,20-40} \sim 3 \times 10^{-15}$~erg~cm$^{-2}$~s$^{-1}$.

To predict {\it NuSTAR} count rates we utilize the Portable, Interactive Multi-Mission Simulator (PIMMS). Considering a purely non-thermal nature for the $(20-40)$~keV X-ray photons, we employ the {\it powerlaw} model in PIMMS, setting the expected un-absorbed flux to $F_{X,20-40}$ and choosing a photon index range of $\delta = (2.5-3)$ \citep{Pillitteri2019}. PIMMS predicts a source count rate in the $(20-40)$~keV band of $10^{-5}$~counts~s$^{-1}$ for both FPMA and FPMB modules when applying a 50\% PSF extraction. \citet{Isola2007} suggest that in powerful flares, the thermal contribution to the $(20-40)$~keV X-ray emission can be significant. However, similar count rates are predicted if we instead assume a purely thermal nature for the $(20-40)$~keV X-ray emission, employing the {\it apec} model with a possible flare temperature range of $kT=(4-8)$~keV \citep{Getman2011}. In the $(20-40)$~keV band, our {\it NuSTAR} data reveal a background count rate of $0.002$~counts~s$^{-1}$. Consequently, not only does the background overwhelms the predicted signal from DQ~Tau in the $(20-40)$~keV band, but it also dominates in any other $>10$~keV band, as clearly demonstrated in Figure~\ref{fig:xray_images}b. But the absence of observed hard ($>10$~keV) X-ray emission from DQ~Tau should not be interpreted as evidence against the applicability of the thick-target model to DQ~Tau's flares.

\subsection{Characteristic X-ray Emission} \label{sec:discussion_characteristic_xrays}

\citet{Wolk05,Favata2005,Getman08a} have demonstrated that very young stars ($t<5$~Myr) spend approximately three-fourths of their time in a quasi-constant characteristic X-ray level, which is likely a result of the superposition of unresolved micro-flaring and nano-flaring \citep{Aschwanden2000b}. Our observations, utilizing data from {\it Chandra}, {\it Swift}, and {\it NuSTAR}, as well as archival data from {\it XMM} and {\it Swift}, cover different epochs of DQ~Tau: February 2007 and January 2010 \citep{Getman2011}, March-April 2017 and December 2021 \citep{Getman2022b}, and the current work in July-August 2022. Both this paper and \citet{Getman2022b} establish that the characteristic X-ray level in DQ~Tau remains constant, at $L_X \sim 2 \times 10^{30}$~erg~s$^{-1}$, across the multiple X-ray observations, which are spread out over a time range of 1 to 15 years. In addition, no significant changes in the average surface magnetic field of the primary and secondary binary components are observed within the time period of 2020 to 2022 (Pouilly et al. in prep.).

There have been numerous observations of magnetic dynamo cycles in stars, analogous to the 11-year solar cycle observed on the Sun. These cycles, often referred to as stellar activity cycles, are characterized by long-term periodic variations in magnetic activity indicators, including starspots, photometric variability, chromospheric emission lines, and coronal X-ray emission. Various X-ray studies on stars of different ages have suggested that activity cycles on younger stars may be shorter and less pronounced, if present at all.

For example, \citet{Wargelin2017} conducted X-ray analyses on several mature stars and observed a decrease in the amplitude of quiescent variability as X-ray activity increased. \citet{Coffaro2020,Coffaro2022} discovered that $\epsilon$~Eri, a star approximately 440 million years old, and Kepler-63, a star approximately 210 million years old, exhibited the shortest X-ray cycles and smallest X-ray amplitudes when compared to several older solar-mass stars known to have X-ray cycles. Additionally, their findings suggested that the surfaces of these stars may be extensively (around 60\%-100\%) covered by solar-type X-ray emitting magnetic structures, such as active region cores and flares. Furthermore, \citet{Marino2006} reported no substantial evidence of long-term X-ray variability in the stellar members of the approximately 100 million-year-old open cluster NGC~2516. Similarly, \citet{Maggio2023} reported only a small long-term X-ray variability with an amplitude of approximately $\sim 2$ for the 12 million-year-old young star V1298~Tau.

The absence of evidence for long-term variability in the X-ray characteristic emission of the one million-year-old  DQ~Tau aligns with the notion that younger stars possess larger active regions and more extended X-ray coronal structures \citep{Coffaro2022,Getman22,Getman2023}, which may mitigate the appearance of magnetic dynamo cycling.    


\section{Conclusions} \label{sec:conclusions}

Drawing upon recent observations conducted during a single orbit of DQ~Tau in July-August 2022, which utilized {\it NuSTAR}, {\it Swift}, and {\it Chandra} telescopes (\S~\ref{sec:xray_data_reduction}), alongside previously gathered X-ray and mm-band data from multiple periastrons of DQ~Tau \citep{Salter2010,Getman2011,Getman22}, our study embarks on an extensive analysis to compute the energetic characteristics of X-ray/NUV/optical flares within DQ~Tau (\S~\ref{sec:xray_flare_analyses}).

To provide a broader context, we compare the flare energetics and occurrence rates of DQ Tau with those observed in super-flares from various PMS stars (\S~\ref{sec:comparison_with_coup_mystix_sfincs},\ref{sec:discussion}). Notably, each of the three large X-ray flares identified in 2010, 2021, and 2022, appearing near an orbital phase of 1 (periastron) within the DQ~Tau system, display an exceptional uniformity in X-ray energies. Estimated at approximately $3 \times 10^{35}$~erg, this remarkable consistency implies the presence of a recurring and persistent energy source. We analyze the outcomes of an analytical model pertaining to magnetosphere interaction in eccentric binaries (\S~\ref{sec:xray_flares_near_periastron_discussion}). The model's results align with the injection rate of kinetic energy into the chromosphere by non-thermal electrons, obtained through a combined analysis of mm-band and X-ray flares. Furthermore, the model effectively accounts for a substantial portion of the energies observed in the NUV/optical flares ($20\%$). We have encountered challenges in differentiating between the energetics of periastron events related to magnetic reconnection and those linked to accretion based solely on optical/NUV data. Nevertheless, we have observed the Neupert effect during mm and X-ray flares, and established consistent rates of magnetic energy release and non-thermal electron injection.
Additionally, the optical-to-X-ray energy ratios between DQ~Tau and prominent solar/stellar flares demonstrate alignment when we utilize optical energy levels predicted by the model of colliding magnetospheres.Moreover, two distinct optical/NUV peaks precede their corresponding X-ray counterparts, and both the optical and X-ray events, influenced by sustained heating from colliding magnetospheres, display similar durations. These collective findings substantiate the notion that the mm/X-ray periastron flares, and potentially, the magnetic-related components of the optical/NUV emissions, adhere to the classical solar/stellar non-thermal thick-target model (\S~\ref{sec:xray_flares_near_periastron_discussion}). {\it NuSTAR} observations encountered high background levels, impeding the detection of anticipated non-thermal hard X-rays (\S~\ref{sec:discussion_non_thermal_xrays}). 

Serendipitously, we discovered X-ray super-flares outside of periastron, potentially related to interacting magnetospheres (\S~\ref{sec:xray_flares_away_from_periastron_discussion}). 

The absence of evidence for long-term variability in the baseline X-ray emission of $\sim 1$~Myr old DQ~Tau is consistent with the understanding that younger stars typically exhibit larger active regions and more extensive X-ray coronal structures. This may contribute to the reduction of observable magnetic dynamo cycling (\S~\ref{sec:discussion_characteristic_xrays}).

\section{Acknowledgments}
We thank the referee for their time and many very useful comments that improved this work. We thank Vitaly Akimkin (Institute of Astronomy) and Fred Adams (University of Michigan) for useful discussions. The authors express their gratitude to the {\it Chandra} science team for enabling the DDT observations conducted in this study. This project is supported by the {\it NuSTAR} NASA grant  80NSSC22K1822 (K. Getman, Principal Investigator) and the {\it Chandra} ACIS Team contract SV4-74018 (G. Garmire \& E. Feigelson, Principal Investigators), issued by the {\it Chandra} X-ray Center, which is operated by the Smithsonian Astrophysical Observatory for and on behalf of NASA under contract NAS8-03060. The {\it Chandra} Guaranteed Time Observations (GTO) data used here and listed in \citet{Getman05,Getman2011,Getman2021} were selected by the ACIS Instrument Principal Investigator, Gordon P. Garmire, of the Huntingdon Institute for X-ray Astronomy, LLC, which is under contract to the Smithsonian Astrophysical Observatory; contract SV2-82024. This paper employs a list of new {\it Chandra} datasets, obtained by the {\it Chandra} X-ray Observatory, contained in~\dataset[DOI: 10.25574/cdc.164]{https://doi.org/10.25574/cdc.164}.

https://doi.org/10.25574/cdc.164

\vspace{5mm}
\facilities{NuSTAR, CXO, Swift}

\software{R \citep{RCoreTeam20}, HEASOFT, CIAO}



\clearpage\newpage

\bibliography{my_bibliography}{}
\bibliographystyle{aasjournal}

\end{document}